\documentclass{article}
\usepackage{amsthm}
\usepackage{amsmath}
\usepackage{amstext}
\usepackage{amsfonts}
\usepackage{amssymb}
\usepackage{amsopn}
\usepackage{graphicx}
\usepackage{cite}
\usepackage{bm,bbm}
\usepackage{epsfig,epic}
\usepackage{subfigure}
\usepackage{stackrel}

\usepackage[left=3cm, right=3cm, top=3cm, bottom=3cm]{geometry}

\renewcommand{\phi}{\varphi}
\newcommand{\iy}{\infty}
\renewcommand{\leq}{\leqslant}
\renewcommand{\geq}{\geqslant}

\newcommand{\ket}[1]{| #1 \rangle}
\newcommand{\ketbra}[2]{| #1 \rangle \langle #2 |}

\DeclareMathOperator{\trace}{Tr}
\DeclareMathOperator{\I}{I}
\DeclareMathOperator{\id}{id}
\DeclareMathOperator{\Wg}{Wg}
\DeclareMathOperator{\Mob}{Mob}

\DeclareMathOperator{\argmin}{argmin}
\DeclareMathOperator{\rk}{rk}
\DeclareMathOperator{\Span}{Span}
\newcommand{\N}{\mathbb{N}}

\newcommand{\R}{\mathbb{R}}
\newcommand{\C}{\mathbb{C}}
\newcommand{\E}{\mathbb{E}}

\renewcommand{\H}{\mathcal{H}}

\newcommand{\U}{\mathcal{U}}
\newcommand{\D}{\mathcal{D}}
\newcommand{\M}{\mathcal{M}}

\renewcommand{\S}{\mathcal{S}}

\newcommand{\ol}{\overline}
\newcommand{\isom}{\simeq}
\DeclareMathOperator{\Rem}{Rem}

\newcommand{\Pvertex}{\Pi_\text{vertex}}
\newcommand{\Ptrace}{\Pi_\text{trace}}
\newcommand{\Pedge}{\Pi_\text{edge}}
\newcommand{\Pentangle}{\Pi_\text{entangle}}

\newtheorem{theorem}{Theorem}[section]
\newtheorem{definition}[theorem]{Definition}
\newtheorem{proposition}[theorem]{Proposition}
\newtheorem{remark}[theorem]{Remark}
\newtheorem{lemma}[theorem]{Lemma}

\begin{document}

\begin{center}
{\Large  {Random graph states, maximal flow \\ and Fuss-Catalan distributions}}

\vskip 0.5cm
  {\bf Beno\^it Collins$^{1,2}$, Ion Nechita$^1$ and Karol {\.Z}yczkowski $^{3,4}$}

\vskip 0.3cm
$^1$Department of Mathematics and Statistics, University of Ottawa, ON K1N8M2, Canada
 
$^2$CNRS, Universit\'e Claude Bernard Lyon 1, Institut Camille Jordan, France.
  
$^3$Institute of Physics,   Jagiellonian University, 
 ul. Reymonta 4, 30-059 Krak{\'o}w, Poland

$^4$Center for Theoretical Physics, PAS, 
         Al. Lotnik{\'o}w 32/44, 02-668 Warszawa, Poland

\vskip 0.5cm

\begin{abstract}
For any graph consisting of $k$ vertices and $m$ edges we construct an ensemble of random pure quantum states which describe a system composed of $2m$ subsystems. Each edge of the graph represents a bi-partite, maximally entangled state. Each vertex represents a random unitary matrix generated according to the Haar measure, which describes the coupling between subsystems. Dividing all subsystems into two parts, one may study entanglement with respect to this partition. A general technique to derive an expression for the average entanglement entropy of random pure states associated to a given graph is presented. Our technique relies on Weingarten calculus and flow problems. We analyze statistical properties of spectra of such random density matrices and show for which cases they are described by the free Poissonian (Marchenko-Pastur) distribution. We derive a discrete family of generalized, Fuss-Catalan distributions and explicitly construct graphs which lead to ensembles of random states characterized by these novel distributions of eigenvalues.
\end{abstract}

\end{center}
\section{Introduction} 

The phenomenon of quantum entanglement in physical systems 
remains a subject of considerable scientific interest. The case of
entanglement in  bi-partite systems is relatively well understood \cite{HHHH09},
but the case of systems consisting of several subsystems is much
more demanding and complicated. For instance, 
the measures characterizing quantitatively quantum entanglement
worked out for the simplest case of two subsystems \cite{MCKB05,PV07}
are often not capable to describe the complexity
of multi-partite entanglement.
On the other hand, the measures applicable in this case,
like the geometric measure of entanglement \cite{WG03}
related to the minimal distance of the analyzed pure state
to the set of separable states, are in general not easy to compute.

Not being in position to characterize entanglement 
of a concrete state of a composed quantum system, 
one often tries to describe properties of a `generic' quantum state.
To this end one constructs {\sl ensembles} of quantum states
and computes interesting quantities averaged over the entire ensemble.

In the simplest case of a system consisting of two subsystems
one considers an ensemble
of random pure states distributed uniformly with respect to the natural,
unitarily invariant (Fubini-Study) measure.
Quantum entanglement with respect to the partition into these
subsystems  called $A$ and $B$
can be described by its {\sl entanglement entropy},
$E(\phi)=S(\trace_B |\phi\rangle \langle \phi|)$.
Here $S(\sigma)=-\log \trace \sigma \log \sigma$
denotes the von Neumann entropy of the state $\sigma$
obtained by the partial trace of the state 
$|\phi\rangle \langle \phi|$. As the initial state is pure,
the resulting entropy does not depend on the subsystem, 
with respect to which the partial trace is taken.
The entropy is bounded from above by the log of the smaller dimensionality,
$S_{\rm max}= \log d_m$, where $d_m=\min \{d_A,d_B\}$.
If this bound is saturated, the initial pure state
is called {\sl maximally entangled}.

The mean entanglement entropy $\langle E(\phi)\rangle_{\phi}$,
averaged over the ensemble of random pure states
depends only on the dimensionality of both subsystems.
An explicit formula for the  mean entropy of a subsystem
was first conjectured by Page \cite{Pa93} and later proved in \cite{FK94}.
Due to the measure concentration phenomenon
the entanglement entropy of a generic pure state is close
to the mean value $\langle E\rangle$. 
Concrete bounds relying on the Levy's lemma \cite{HLW06}
specify the probability to find a random state 
with entropy of entanglement smaller than $\langle E\rangle$ by $\epsilon$.

Note that the Fubini-Study measure on the
space of  pure states of the entire system followed
by the partial trace over subsystem $B$
induces a probability measure in the space
of mixed quantum states describing the subsystem $A$.
These induced measures \cite{Br96,ZS01}
are parameterized by the dimensionality of the auxiliary subsystem $B$.

Analyzing an ensemble of the reduced random states $\sigma$, 
Page derived the probability distribution for its eigenvalues.
Up to an overall normalization constant this problem
is equivalent to finding the density of the spectrum
of the Wishart matrices $W=GG^{*}$ 
(where $G$ is a rectangular complex random matrix of the Ginibre ensemble)
studied  earlier by Marchenko and Pastur \cite{MP67}.
This very distribution depending only on the ratio
$c=d_B/d_A$  of dimensions of both subsystems,
is also called {\sl free Poissonian},
as it corresponds to the free convolution of random matrices
\cite{BLS96,nica-speicher}. 

In the general case of pure random quantum states describing
multipartite systems one usually studies entanglement with respect to
various cuts of the entire system into two parts \cite{KZM02,Sc04,FFMPP09},
so the standard measures of the bi-partite entanglement can be applied.
Statistical properties of random pure states change
if one breaks the overall unitary invariance, characteristic to the Fubini-Study
measure, and introduces some structure into the multi-partite system.
In particular, for certain numbers of subsystems there exist perfect
{\sl  maximally multipartite entangled states} (MMES),
such that they are maximally entangled with respect to any
bi-partition \cite{Sc04,FFPP08}.

Random quantum states are useful to tackle 
various problems of theoretical physics.
A key conjecture of the theory of quantum information processing
concerning the additivity of minimal output entropy 
was recently shown to be false \cite{Ha09,BH09}.
It is worth to emphasize that the original reasoning was not constructive
but it was based on relations between the average quantities 
computed with respect to an ensemble of quantum states.
Entanglement between random states is also interesting 
in systems motivated by condensed matter physics \cite{LR09,GCPP09}.
 For a discussion of a relationship between quantum criticality,  Anderson transition and
the average entanglement entropy see the recent 
review \cite{RM09} and references therein.

The theory of random pure states proved to be useful 
to analyze the information flow and the entropy of black holes.
The model of Page \cite{Pa93b} consisted of two subsystems, the dimensions 
of which served as parameters of the model.
In later  models  one studied pure states of a system composed
of four parties \cite{HP07, BSZ09}.
The system contains two pairs of maximally entangled states
which relate subsystems $A,A'$ and $B,B'$ respectively.
The subsystems $A$ and $B$ are directly coupled together,
and the action of an unknown Hamiltonian is mimicked
by a random unitary matrix distributed according to the Haar measure.
Such a system can symbolically be depicted by a graph containing two edges,
which represent two maximally entangled states 
and a vertex denoting a random unitary matrix.
This very situation is shown in Fig. \ref{fig:2-edge-graph}b,
where a slightly different notation is used.

The main aim of this paper is to extend this construction for an arbitrary
 (undirected) graph and to investigate properties of the resulting ensembles
of multi-partite random quantum pure states.
Any graph with $m$ edges will be associated with a quantum system
consisting of $2m$ subsystems. Each edge corresponds to a maximally
entangled state while each vertex represents a random unitary matrix.
The dimensions of two subsystems linked by a given edge of the graph
are assumed to be equal, but besides this constraint the dimensions
of all the subspaces can be treated as free parameters of the model.
As the physical interaction is modeled by a set of random
unitary matrices each graph defines an entire ensemble, 
we call them {\sl random graphs states}.

Note that such an {\sl ensemble} of quantum random states associated with
a given graph differs form the deterministic construction of {\sl graph states}
introduced by Hein et al. \cite{HEB04}.
Furthermore, our approach is not related to {\sl quantum graphs}
reviewed by Gnutzmann and  Smilansky \cite{GS06},
which describe quantum particles (or waves) traveling along a graph.
A graphic representation of ensembles of random states analyzed in this work
looks slightly similar to the one used to define 
the Projected Entangled Pair States (PEPS) formalism, 
capable to describe complex many-body systems \cite{VC04,VWPC06}.
While in the latter setup  the subsystems entangled with auxiliary systems
are coupled together by a projection on a low dimensional subspace,
in our approach such a coupling is described by a random unitary matrix.

Any graph defines the topology of
couplings between the physical subsystems.
Choosing a certain set of the subsystems one can average
the pure quantum state $|\psi\rangle$ over the remaining subsystems.
Technically one performs the partial trace the auxiliary subsystems,
which leads to a reduced state which generically is not pure.
Hence selecting a concrete graph and specifying the subsystems 
pertaining to the environment we define an ensemble of random mixed states.

In this work we develop general techniques suitable to describe
spectral properties of random density matrices associated with a graph.
Our technique relies on Weingarten calculus \cite{collins-imrn, collins-nechita-1}, 
and the observation that its asymptotics boil down to a min-flow problem \cite{cormen}. 
In the asymptotic limit, as the dimension of the quantum states 
goes to infinity, statistical properties the spectrum of a random state can be established,
so explicit formulae for its purity and entropy are derived.

We identify these ensembles of random states for which spectra 
are described by the free Poisson (Marchenko-Pastur) distribution.
In some other cases the spectra are described by a discrete family of probability distributions $\pi^{(s)}$,
which can be be considered as generalizations of the free Poissonian
distribution. We call them {\sl Fuss-
Catalan distributions} (FC),
since their moments are related to the Fuss-Catalan numbers,
known in the combinatorics and 
free probability calculus \cite{armstrong, banica-etal}.

The Fuss-Catalan probability distributions are also closely related
with properties of {\sl products} of non-hermitian random matrices.
In general, products of random matrices are often studied
in context of various problems of statistical physics  \cite{CPV93}.
Recent studies on products of Ginibre matrices
concern multiplicative diffusion processes \cite{GJJN03},
rectangular correlation matrices used in macroeconomic
time series \cite{BLMM07} and lattice gauge field theories \cite{LNW08}.
Spectral properties of a product of random Ginibre matrices
were  analyzed  in a recent paper of by Burda et al. \cite{BJW10}.

In the simplest case $s=2$ random matrices described 
by the distribution $\pi^{(2)}$ have the structure
$G_2G_1G_1^{*}G_2^{*}$
where $G_1$ and $G_2$ are independent
random Ginibre matrices.
In the general case of an arbitrary integer $s$
the random states are proportional
to the Wishart matrix $W=GG^{*}$,
constructed out of a product of $s$ independent Ginibre matrices, 
$G= \prod_{i=1}^s G_i$.

To analyze statistical properties of random states
whose definition is based on random unitary matrices,
one needs to perform averages over the group of unitary matrices.
Instead of using explicit formulae derived by Mello \cite{Me90}
we found it more convenient to use the Weingarten calculus
(see e.g. \cite{collins-imrn,collins-sniady}).
In order to evaluate expectation values for random tensors 
we are going to use  a diagramatic approach and a graphical calculus 
recently developed in \cite{collins-nechita-1}.

This paper is organized as follows: ensembles of pure states corresponding to 
classical graphs are defined in section \ref{sec:graph-states}.
In section \ref{sec:marginals}, we discuss mixed states
which arise by taking the partial trace over a specified
set of subsystems.
In section \ref{sec:unitary-integration} graphical and combinatorial tools
used to perform integration over the unitary group
are described.  The main result of this work --- a general technique
to compute the moments of the spectrum 
of ensembles of mixed states obtained by 
partial trace --- is presented in section \ref{sec:main-result}.
In section \ref{sec:applications} we analyze ensembles associated to
some particular graphs (star graphs, cycle graphs) 
and find ensembles of mixed states 
characterized by Marchenko-Pastur and Fuss-Catalan
distributions. Some other graphs which lead to other, ``exotic'' distributions are also provided.
The paper is concluded in section 7 in which we summarize results obtained
and present a list of open questions.

\section{Classical graphs, quantum pure states and random matrices}\label{sec:graph-states}

In this section we describe a family of quantum pure states 
that are associated in a natural way with undirected graphs.
 The underlying idea is that vertices of graphs correspond to quantum
 systems and edges describe the entanglement between the systems.
 A particular feature of the states we consider is that the entanglement 
described by the edges will be maximal.

\begin{figure}[ht]

\centering
\subfigure[]{\includegraphics{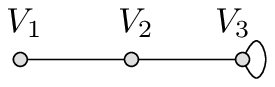}}\quad\quad
\subfigure[]{\includegraphics{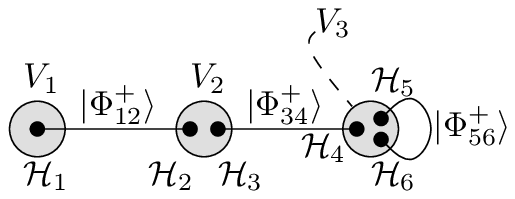}}
\caption{(a) Graph consisting of $3$ vertices and $3$ bonds,
one of which is connected to the same vertex so it forms a loop; (b) the corresponding ensemble of random pure states
defined in a Hilbert space composed of $6$ subspaces
represented by dark dots.
Each bond corresponds to a maximally entangled state spanned
on two subspaces connected by the bond,
while each vertex corresponds to a random unitary matrix.}
\label{fig:graph-example}
\end{figure}

Consider an undirected graph $\Gamma$ consisting of $m$ \emph{edges} (or \emph{bonds})
 $B_1, \ldots, B_m$ and $k$ \emph{vertices} $V_1, \ldots V_k$. We allow multiple edges
 between two vertices, as well as vertex loops.
Note that the graphs considered here are not metric,
so we do not discuss the lengths of the edges.
 Let $b_i$ denote the
degree of the vertex $V_i$ defined as the number of edges attached
 to this vertex (each loop counts twice). 
As an example, for the graph $\Gamma$ in Figure \ref{fig:graph-example}, we have $b_1 = 1$, $b_2 = 2$, $b_3 = 3$. 

We are going to analyze quantum states belonging to the Hilbert
space with the $n=2m$-fold tensor product structure, 
${\H}={\H}_1 \otimes \cdots \otimes {\cal H}_{2m}$.
The dimension of the subspace  ${\cal H}_{i}$ 
is $d_i N$, where the parameters $d_i>0$ are fixed 
while the number $N$ is arbitrary and
we will eventually discuss the limit $N\to \infty$.
The total dimension of space ${\cal H}$
reads thus $D=\dim \H = \left(\prod_{j=1}^{n}d_j \right) N^n$.

A graph $\Gamma$ with $m$ edges induces a tensor product structure of the
 $2m$-fold tensor product space $\H$. The vertices $V_l$ of $\Gamma$ induce a partition
 $\Pvertex$ of the set $[n] = \{1, 2, \ldots, 2m\}$, with blocks $C_1, \ldots, C_k$. The reader can think
 about the Hilbert spaces $\{\H_j \; | \; j \in C_l\}$ living inside the vertex $V_l$.
 For each vertex $V_l$, we introduce the Hilbert space $W_l = \otimes_{j \in C_l} \H_j$
which describes the $b_l$ subsystems of the vertex $V_l$.

Each edge of the graph represents the maximally entangled 
states between two Hilbert spaces $\H_i$ and $\H_j$. 
More precisely, we shall ask that $\dim \H_i = \dim \H_j = d_i N$ and we put
\begin{equation}
 \ket{\Phi^+_{ij}} =  \frac{1}{\sqrt{d_i N}} \sum_{x=1}^{d_i N} \ket{e_x}\otimes \ket{f_x},
\label{eq:Bell}
\end{equation}
where $\ket{e_x}$ and $\ket{f_x}$ are orthonormal bases for $\H_i$ and $\H_j$. The actual choice of the two bases is not important in what follows, but we shall assume that these bases are fixed.

In order to introduce an  ensemble of \emph{random graph states}
belonging to $\H$
we are going to use $k$ independent, random matrices $U_1, U_2, \ldots, U_k$,
distributed according to the Haar measure.
The matrix  $U_l$ corresponds to the $l$-th vertex of the graph, $V_l$.
It acts on the space $W_l$ (and thus on  ${b_l}$ subspaces) and
describes the coupling at a given vertex.
To specify the subspaces coupled by a unitary matrix 
it will be  also convenient to use an alternative notation 
and write $U_l=U_{i_1,\ldots,i_p}$ where $p=b_l$ denotes the
degree of the vertex. See Figure \ref{fig:graph-example} for a concrete example
of a graph and a graph ensemble.

\begin{figure}[ht]
\centering
\subfigure[]{\includegraphics{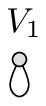}}\quad\quad
\subfigure[]{\includegraphics{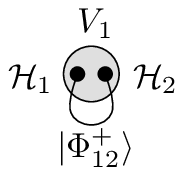}}\quad\quad
\subfigure[]{\includegraphics{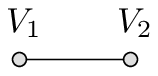}}\quad\quad
\subfigure[]{\includegraphics{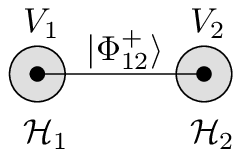}}
\caption{Graphs with one edge: a loop on one vertex, in simplified notation (a) and in the standard notation (b), and two vertices connected by one edge, in simplified notation (c) and in the standard notation (d).}
\label{fig:1-edge-graph}
\end{figure}

The simplest graph ensembles described by our model have only one edge. We consider two cases, one where the unique edge is a loop of one vertex, and the second one where the edge connects two vertices (see Figure \ref{fig:1-edge-graph}).

In the first situation, the graph has a single vertex of degree two.  This graph corresponds to 
the system consisting of $2m=2$ subsystems. The edge represents the maximally entangled state, 
but due to the global unitary matrix $U_1=U_{1,2}$, 
 $\ket \Psi=U_{1,2}\ket{\Phi_{12}^+}$ is 
 a random state in $\H_1 \otimes \H_2$.
Since $U_{1,2}$ is generated according to the Haar measure,
the random state $\ket \Psi$ is distributed 
uniformly with respect to the Fubini-Study measure,
as in the model of  Page \cite{Pa93},
so in this case the bond forming the loop 
is in a sense redundant.

In the linear case, we have again only one edge, $m=1$, 
two vertices, $k=2$, and each vertex contains only one node. 
The graph is represented in Figure \ref{fig:1-edge-graph} (c) and (d), both in the simplified and in the
standard form.
The random quantum pure state associated 
to this graph is a generic maximally entangled state, 
\begin{equation}
\ket \Psi = \left[ U_1 \otimes U_2 \right]\frac{1}{\sqrt{dN}} \sum_{i=1}^{dN} \ket{e_i} \otimes \ket{f_i} = \frac{1}{\sqrt{dN}} \sum_{i=1}^{dN} \ket{U_1 e_i} \otimes \ket{U_2 f_i},
\label{eq:Bell2}
\end{equation}
where we set $d_1 = d_2 = d$ and where $U_1 , U_2$ represent  independent random unitary matrices. Here $\{e_i\}_i$ , $\{f_i\}_i$ denote some fixed bases of $\C^{dN}$.

\begin{figure}[ht]
\centering
\subfigure[]{\includegraphics{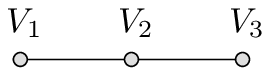}}\quad\quad
\subfigure[]{\includegraphics{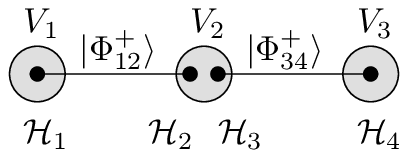}}
\caption{A linear $2$-edge graph, in the simplified notation (a) and in the standard notation (b).}
\label{fig:2-edge-graph}
\end{figure}

As a third and final example, the next simplest graph one could imagine has a linear shape, $m=2$ edges, $k=3$ vertices $V_1$, $V_2$, $V_3$ and $n=2m=4$ nodes, as in Fig. \ref{fig:2-edge-graph}.

In general, any graph $\Gamma$ is associated with 
a collection of $n=2m$ subspaces ${\cal H}_1 ,  \ldots, {\cal H}_{2m}$,
each endowed with some fixed orthonormal basis.
We consider two partitions of the set $[n] = \{1, 2, \ldots, n\}$.
A pair partition $\Pedge$ with the property that $\forall \{i,j\} \in \Pedge, \; d_i = d_j$,
represents all bonds in the graph and thus the
$m$ entangled states.
The second partition $\Pvertex$ consists of $k$ blocks of size $b_i$,
which represent the vertices of the graph,
and encode the random unitary operators $U_i$.
For instance, in the case of $n=6$ subspaces presented in Fig.\ref{fig:graph-example} the two partitions read $\Pedge=\{\{1,2\}, \{3,4\}, \{5,6\}\}$
and  $\Pvertex=\{ \{1\}, \{2,3\}, \{4,5,6\}\}$.

These two partitions allows us to introduce a general definition of an
{\sl ensemble of graph random states}

\begin{equation}
 \bigotimes_{i=1}^n {\cal H}_i \ \ni \ \ket{\Psi_\Gamma} \; =\;  \left[ \bigotimes_{C \in \Pvertex} 
U_C \right]\left( \bigotimes_{\{i,j\} \in \Pedge} 
\ket{\Phi^+_{i,j}}
 \right) ,
\label{eq:def-graph-state}
\end{equation}
where $U_C \in \U(W_C) = \U(\otimes_{i \in C} {\cal H}_i)$ are Haar 
independent random unitaries and $\ket{\Phi^+_{i,j}}$ are the
maximally entangled states (\ref{eq:Bell})
defined with respect to the fixed orthonormal bases 
of ${\cal H}_i$ and ${\cal H}_j$.

\section{Marginals of random graph states}
\label{sec:marginals}

To study non-local properties of the random graph state $\ket \Psi$ associated to a graph $\Gamma$
it is useful to specify a partition of the set of all $2m$
subsystems into two groups, $\Ptrace=\{S,T\}$.
Then it is possible to treat the multi-partite state $\ket \Psi$ 
as if it where bipartite and to analyze its entanglement with respect to this concrete partition.
The entanglement of  a  bipartite pure state can be measured
by the von Neumann entropy of the reduced state
obtained from the initial projector $|\Psi \rangle \langle \Psi|$ 
by the partial trace.
We will analyze the partial trace of the pure state 
$\ketbra{\Psi}{\Psi}$ over a subspace ${\cal H}_T$
defined by the subset $T$ of the set $[n]$.
Then $S$ denotes the complementary subset,
so the sum of their elements $|T|+|S|$ is equal to $n$
and the entire Hilbert space can be written as 
a tensor product, ${\cal H}={\cal H}_T \otimes {\cal H}_S$.
Such a partial trace will be briefly denoted by $\trace_T$
and the resulting density matrix is supported then
by the subspace  $\ H_S$,

\begin{equation}\label{eq:parttrace}
\rho_S = \trace_{T} \ketbra{\Psi}{\Psi}.
\end{equation}

\begin{figure}[htbp]
\centering
\includegraphics{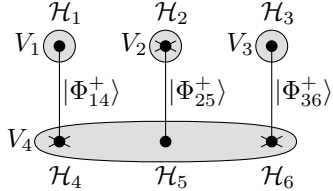}
\caption{Exemplary graph representing random states
  supported on $n=6$ subspaces and defined by entangled states
$\ket{\Phi^+_{14}}$, $\ket{\Phi^+_{25}}$ and 
$\ket{\Phi^+_{36}}$ and a non-local random unitary matrix $V_4=U_{4,5,6}$
of dimension $d_4 d_5 d_6 N^3$ and local unitary matrices 
$V_1, V_2$ and $V_3$.
Partial trace of this state over the subspace ${\cal H}_T$ 
defined by the set $T=\{2,4,6\}$, represented by crosses,
provides the reduced state $\rho_S$
supported on subspaces corresponding to the set $S=\{1,3,5\}$
and represented by full (non crossed) dots.
}
\label{fig:broadcasting}
\end{figure}

Graphically, partial traces are denoted
at the graph   by ``crossing'' the spaces $\H_i$ which are being traced out.
For example, in Figure \ref{fig:broadcasting}, the partition defining the bi-partite structure is $\Ptrace = \{S = \{1,3,5\}, T=\{2,4,6\}\}$. In the diagram, the dots representing the subspace $\H_2$, $\H_4$ and $\H_6$ are ``crossed'', and the partial trace is taken on the tensor product of those spaces $\H_T = \H_2 \otimes \H_4 \otimes \H_6$.

Before exploring the quantitative entanglement of $\ketbra{\Psi}{\Psi}$, we can make some qualitative remarks right away. Looking at Figure \ref{fig:broadcasting} we see that 
the systems $\H_1$ and $\H_3$ are not entangled
before  the  random unitary transformation $V_4=U_{4,5,6}$ was applied. Indeed, before applying the random unitary matrices, the state of the system reads
\begin{equation}
\ket{\tilde \Psi} = \ket{\Phi^+_{14}} \otimes	\ket{\Phi^+_{25}} \otimes \ket{\Phi^+_{36}}.
\end{equation}
The product structure of the state $\ket{\tilde \Psi}$ is given by the supremum of the vertex and edge partition, $\Pvertex \vee \Pedge$. However, after the unitary transformations, subsystems $\H_1$ and $\H_3$ become entangled (generically); one might say that entanglement was broadcasted by the unitary matrix $V_4$.

In general, we define the partition $\Pentangle$ by
\begin{equation}
 i \stackrel{\Pentangle}{\sim} j \iff i \stackrel{\Pvertex}{\sim} j \quad \text{ and } \quad [i,j]_{\Pvertex} \nsubseteq T,
\end{equation}
where $[i,j]_{\Pvertex}$ denotes the block of $i$ and $j$ in $\Pvertex$.
In other words, $\Pentangle$ is obtained from $\Pvertex$ in the following way: 
if a block of $\Pvertex$ is \emph{entirely} traced out, then replace it by singletons,
 otherwise keep it as it is.
 The block structure of $\rho_S$ is then given by the restriction of the 
partition $\Pentangle \vee \Pedge$ to the set $S$. For the graph in Figure \ref{fig:broadcasting}, we have $n=6$, $\Pedge=\{\{1,4\}, \{2,5\}, \{3,6\}\}$, $\Pvertex=\{ \{1\}, \{2\}, \{3\}, \{4,5,6\}\}$ and $S=\{1, 3, 5\}$. 
Since the unitary $U_{4,5,6}$ is not completely traced out, it is capable to broadcast the entanglement between subsystems $1$ and $3$.

We finish this section by a remark which can simplify the structure of graph state marginals in some cases. If a marginal of the pure state (\ref{eq:def-graph-state}) 
is considered, the  unitaries $U_C$ corresponding to singletons $C = \{i\} \in \Pvertex$ 
in equation (\ref{eq:def-graph-state}) can be replaced by the identity operator if the pair $j$ of $i$ 
in $\Pedge$ is not a singleton in $\Pvertex$. This follows from the invariance of the Haar measure on the unitary group and from the fact that one can extract an independent unitary from the block of $j$ in $\Pvertex$.

\section{Graphical and combinatorial tools for unitary integration}\label{sec:unitary-integration}

\subsection{Permutations, non-crossing partitions}
\label{sec:notations}

In this section,
we introduce notation that will be used in the entire paper and which may be non-standard. For an integer $n$, let $[n]$ denote a set of $n$ elements $\{1, 2, \ldots, n\}$. For a collection of numbers $\{d_i\}_i$, Hilbert spaces $\{\H_i\}_i$ and for a set of indices $I$, we are going to use the notation  ${\cal H}_I = \otimes_{i \in I} {\cal H}_i$ and  $d_I = \prod_{i \in I} d_i$. 

We denote by $\S_q$ the group of permutations on $q$ elements. For a permutation $\sigma$, we call $|\sigma|$ its length, that is the minimum number of transpositions necessary to obtain $\sigma$ and $\#\sigma$ the number of disjoint cycles of $\sigma$. One has $|\sigma| + \#\sigma = q$ for all permutations $\sigma \in \S_q$. Note that the absolute value is also used in this work to denote the cardinality of a set $|A|$.

We consider the lattice $\mathcal{P}_q$ of partitions of $[q]=\{1,\ldots ,q\}$, and
say that $\Pi \leq \Pi'$ if, for any block $V$ of $\Pi$ there exists a block $V'$
of $\Pi'$ that contains $V$, $V \subseteq V'$. For any \index{$\vee$} \index{$\wedge$}
partitions $\Pi$ and $\Pi'$ define $\Pi\vee\Pi'$ (resp. $\Pi\wedge\Pi'$) to be
the least upper bound (resp. the greatest lower bound) of $\Pi$ and $\Pi'$ with respect to the previously defined partial order, and also let $\hat 1_q=\{\{1,\ldots ,q\}\}$ \index{$0_q$}\index{$1_q$}
(resp. $\hat 0_q=\{\{ 1\},\ldots ,\{ q\}\}$) the greatest (resp. smallest) element in $\mathcal P_q$.

The permutation group $S_q$ admits a natural left action
on the partitions of $[q]$. Call 
$\mathcal{O}_{\Pi}$ the orbit to which $\Pi$ belongs. These orbits
are in natural one to one correspondence with the partitions of the
integer $q$.
For any permutation $\sigma\in S_q$, we denote by  $[\sigma]$ the partition of $[q]$ whose
blocks are the orbits of $\sigma$.

The lattice $\mathcal{P}_q$ admits a subset of interest, of set of all \emph{non-crossing partitions}.
We denote it by $NC(q)$. A partition is called non-crossing iff there exists no two different blocks $X,Y$
of the partition and $a,c\in X$, $b,d\in Y$, with $a<b<c<d$.
$NC(q)$ is a sub-poset of $\mathcal{P}_q$ and it turns out to be a lattice as well. 
The operation $\wedge$ is  the same for
$NC(q)$ and $\mathcal{P}_q$, however the supremum operation $\vee$ is different in the non-crossing setting, as it can be seen by the example $\Pi=\{\{1\},\{3\},\{2,4\}\}$ and $\Pi' = \{\{1,3\},\{2\},\{4\}\}$.

We consider the vector subspace $A_p$ of functions $\mathbb{C}^{NC(q)\times NC(q)}$
whose value is $0$ when $\Pi_1\leq \Pi_2$ does not hold.
This vector space is endowed with a convolution operation $*$
\begin{equation}
\label{eq:convolution}
[f_1*f_2](\Pi_1 ,\Pi_2):=\sum_{\Pi_3 \text{ s.t. } \Pi_1\leq\Pi_3\leq\Pi_2}f_1(\Pi_1,\Pi_3)\; f_2(\Pi_3,\Pi_2).
\end{equation}
This turns $A_p$ into an algebra with unit $\delta_{\Pi_1,\Pi_2}$.
For each $\Pi_1,\Pi_2\in\mathcal{P}$ such that $\Pi_1\leq\Pi_2$, 
there exists a partition $\Pi$ such that
the interval $[\Pi_1 ,\Pi_2]$ is isomorphic as a lattice to the lattice
$[0_q,\Pi]$. If $\mathcal{O}_{\Pi}=(p_1,\ldots,p_k)$ then define the M\"{o}bius function $\Mob$
by
\begin{equation}
\label{eq:mob}
\Mob (\Pi_1 ,\Pi_2)=\prod_{i} \left( (-1)^{i-1}c_{i-1}\right)^{p_i}
\end{equation}
where 
\begin{equation}
c_i=\frac{1}{i+1} \binom{2i}{i} = \frac{(2i)!}{(i+1)! \, i!}	
\end{equation}
 is the $i$-th Catalan number (later we shall see that Catalan numbers are a particular case of a much more general class of combinatorial family, called the Fuss-Catalan family).
We also define the Zeta function $\zeta$ which is equal to one when $\Pi_1 \leq \Pi_2$ and zero otherwise:
\begin{equation}
\zeta (\Pi_1 ,\Pi_2)=1.
\end{equation}
In this paper we will need the fact that the convolution product of the M\"{o}bius 
function $\Mob$ and the Zeta function $\zeta$ is equal to the Kronecker delta 
function $\delta(\Pi_1, \Pi_2) = 1$ iff. $\Pi_1 = \Pi_2$:
\begin{equation}\label{eq:inversion-Mob}
\Mob*\zeta = \delta .
\end{equation}
Note that if we had considered the equation \eqref{eq:convolution} in the general lattice of partitions and had left the definition of $\zeta$
unchanged, we should have modified the definition of $\Mob$ as follows.
\begin{equation*}
\Mob (\Pi_1 ,\Pi_2)=\prod_{i} ((-1)^{i-1}(i-1)!)^{p_i}
\end{equation*}

We finish this section by mentioning the following fact, of crucial importance in what follows: if $\gamma$ is a permutation with one cycle (i.e. a cyclic permutation), the set $\mathcal{N}:=\{ \sigma \in \S_q \; , \; |\gamma\sigma^{-1}|+|\sigma|=|\gamma| = q-1\}$ is in one to one correspondence with the set of non-crossing partitions $NC(q)$. If $\gamma=(1,\ldots ,q)$, which we can always assume up to an overall conjugation of the set $\mathcal{N}$, the correspondence is given by $\sigma \to [\sigma]$. The proof of this result is elementary and belongs to combinatorial folklore.

\subsection{Weingarten calculus}

In this section, we recall a few facts about the Weingarten calculus, 
useful to evaluate averages  with respect to 
the Haar measure on the unitary group.

\begin{definition}
The unitary Weingarten function 
$\Wg(n,\sigma)$
is a function of a dimension parameter $n$ and of a permutation $\sigma$
in the symmetric group $\S_p$. 
It is the inverse of the function $\sigma \mapsto n^{\#  \sigma}$ under the convolution 
for the symmetric group ($\# \sigma$ denotes the number of cycles of the permutation $\sigma$).
\end{definition}

Notice that the  function $\sigma \mapsto n^{\# \sigma}$ is invertible for large $n$, as it
behaves like $n^p\delta_e$ as $n\to\infty$, were $p$ denotes the number of elements
in the permutation group $S_p$. 
We refer to \cite{collins-sniady} for historical references and further details. 
We shall use the shorthand notation $\Wg(\sigma) = \Wg(n, \sigma)$ when the dimension parameter $n$ is obvious.

The following theorem 
relates integrals with respect to the Haar measure on the unitary group $U(n)$
and the Weingarten function $\Wg$. 
(see for example \cite{collins-imrn}):

\begin{theorem}
\label{thm:Wg}
 Let $n$ be a positive integer and
$(i_1,\ldots ,i_p)$, $(i'_1,\ldots ,i'_p)$,
$(j_1,\ldots ,j_p)$, $(j'_1,\ldots ,j'_p)$
be $p$-tuples of positive integers from $\{1, 2, \ldots, n\}$. Then
\begin{multline}
\label{bid} \int_{\U(n)}U_{i_1j_1} \cdots U_{i_pj_p}
\overline{U_{i'_1j'_1}} \cdots
\overline{U_{i'_pj'_p}}\ dU=\\
\sum_{\sigma, \tau\in \S_{p}}\delta_{i_1i'_{\sigma (1)}}\ldots
\delta_{i_p i'_{\sigma (p)}}\delta_{j_1j'_{\tau (1)}}\ldots
\delta_{j_p j'_{\tau (p)}} \Wg (n,\tau\sigma^{-1}).
\end{multline}

If $p\neq p'$ then
\begin{equation} \label{eq:Wg_diff} \int_{\U(n)}U_{i_{1}j_{1}} \cdots
U_{i_{p}j_{p}} \overline{U_{i'_{1}j'_{1}}} \cdots
\overline{U_{i'_{p'}j'_{p'}}}\ dU= 0.
\end{equation}
\end{theorem}

We are interested in the values of the Weingarten function in the limit $n \to \iy$. 
The following result encloses all the data we need for our computations
about the asymptotics of the $\Wg$ function; see \cite{collins-imrn} for a proof.

\begin{theorem}\label{thm:mob} For a permutation $\sigma \in \S_p$, 
let $\mathrm{Cycles}(\sigma)$ denote the set of cycles of $\sigma$. Then
\begin{equation}
\Wg (n,\sigma )=(-1)^{n-\# \sigma}
\prod_{c\in \mathrm{Cycles} (\sigma )}\Wg (n,c)(1+O(n^{-2}))
\end{equation}
and 
\begin{equation}
\Wg (n,(1,\ldots ,d) ) = (-1)^{d-1}c_{d-1}\prod_{-d+1\leq j \leq d-1}(n-j)^{-1}
\end{equation}
where $c_i=\frac{(2i)!}{(i+1)! \, i!}$ is the $i$-th Catalan number.
\end{theorem}

A shorthand for this theorem is the introduction of a M\"{o}bius function 
$\Mob$ on the symmetric
group, invariant under conjugation and multiplicative over the cycles, satisfying
for any permutation $\sigma \in \S_p$:
\begin{equation}
\label{eq:asympt-Wg}
\Wg(n,\sigma) = n^{-(p + |\sigma|)} (\Mob(\sigma) + O(n^{-2})).
\end{equation}
where $|\sigma |=p-\# \sigma $ is the \emph{length} of $\sigma$, i.e. the minimal number of 
transpositions that multiply to $\sigma$, and $\Mob$ was defined at equation \eqref{eq:mob}.
We refer to \cite{collins-sniady} for details about the function $\Mob$.

Next, we recall briefly the results of \cite{collins-nechita-1} for the convenience of the reader and 
in order to make the paper self contained.

\subsection{Axioms of unitary graphical calculus}

The purpose of the graphical calculus introduced in  \cite{collins-nechita-1} is to yield an effective method to evaluate
the expectation of random tensors with respect to the Haar measure on a unitary group. 
The tensors under consideration can be constructed from a few elementary tensors such as the Bell state, 
fixed kets and bras, and random unitary matrices.
In graphical language, a tensor corresponds to a \emph{box}, and an appropriate Hilbertian structure yields a correspondence
between boxes and tensors. 
However, the calculus yielding expectations only relies on diagrammatic operations. 

\begin{figure}[ht]
\centering
\subfigure[]{\label{fig:box}\includegraphics{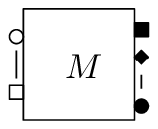}}\qquad\qquad
\subfigure[]{\label{fig:trace}\includegraphics{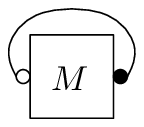}}\qquad\qquad
\subfigure[]{\label{fig:multiplication}\includegraphics{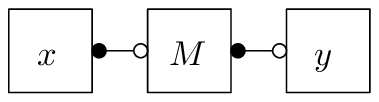}}\\
\subfigure[]{\label{fig:product}\includegraphics{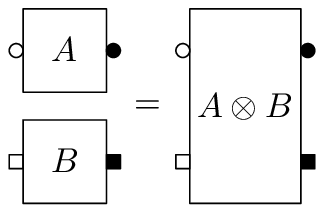}}
\caption{Basic diagrams and axioms: (a) diagram for a general tensor $M$;
 (b) trace of a $(1,1)$-tensor (matrix) $M$; (c) Scalar product $\langle y \;|\; M \;|\; x \rangle$; (d) tensor product of two diagrams}
\end{figure}

Each box $B$ is represented as a rectangle with decorations on its boundary. The decorations are either white or black, and
belong to $S(B)\sqcup S^*(B)$. 
Figure \ref{fig:box} depicts an example of boxes and diagrams.

It is possible to construct new boxes out of old ones by formal algebraic operations such as sums or products.
We call \emph{diagram} a picture consisting in boxes and wires according to the following rule:
a wire may link a white decoration in $S(B)$ to its black counterpart in $ S^*(B)$.
A diagram can be turned into a box by choosing an orientation and a starting point.

Regarding the Hilbertian structure, wires correspond to tensor contractions. 
There exists an involution for boxes and diagrams. It is anti-linear and it turns a decoration 
in $S(B)$ into its counterpart in $ S^*(B)$.
Our conventions are close to those of \cite{coecke,jones}.
They should be familiar to 
 the reader acquainted with existing graphical calculus of various types
(planar algebra theory, Feynman diagrams theory, traced category theory).
Our notations are designed to 
fit well to the problem of computing expectations, as shown in the next section. In Figure \ref{fig:trace}, \ref{fig:multiplication} and \ref{fig:product} we depict the trace of a matrix, multiplication of tensors and the tensor product operation.
For details, we refer to \cite{collins-nechita-1}.

\subsection{Planar expansion}
\label{sec:planar}

The main application of our calculus is to compute expectation of diagrams where some boxes represent random matrices (e.g.
Haar distributed or Gaussian).
For this, we need a concept of \emph{removal} of boxes $U$ and $\ol U$.
A removal $r$ is a way to pair decorations of the $U$ and $\ol U$ boxes appearing in a diagram. 
It therefore consists in  a pairing $\alpha $ of the white decorations of $U$  boxes with the white decorations of $\ol U$ boxes, 
together with a pairing $\beta $ between the black decorations of $U$ boxes and the black decorations of $\ol U$ boxes. 
Assuming that $\D$ contains $p$ boxes of type $U$ and that the boxes $U$ (resp. $\ol U$) are labeled from $1$ to $p$, 
then $r=(\alpha,\beta)$ where $\alpha,\beta$ are permutations of $\mathcal{S}_p$.

Given a removal $r \in \Rem(\D)$, we construct a new diagram $\D_r$ associated to $r$, which has the important property that it no longer contains boxes of type $U$ or $\ol U$. 
One starts by erasing the boxes $U$ and $\ol U$ but keeps the decorations attached to them. 
Assuming that one has labeled the erased boxes $U$ and $\ol U$ with integers from $\{1, \ldots, p\}$, one connects \emph{all} the (inner parts of the) \emph{white} decorations of the $i$-th erased $U$ box with the corresponding (inner parts of the) \emph{white} decorations of the $\alpha(i)$-th erased $\ol U$ box. In a similar manner, one uses the permutation $\beta$ to connect black decorations. 

In \cite{collins-nechita-1}, we proved the following result:
\begin{theorem}\label{thm:Wg_diag}
The following holds true:
\[\E_U(\D)=\sum_{r=(\alpha, \beta) \in \Rem_U(\D)} \D_r \Wg (n, \alpha\beta^{-1}).\]
\end{theorem}

\section{Main result: computing the moments of $\rho_S$ and a flow problem}\label{sec:main-result}
 
\subsection{A formula for the moments of $\rho_S$}\label{sec:recipe}

This section contains one of the main results of the paper, Theorem \ref{thm:moments}. 

We present a general method for computing the moments of the random density matrix $\rho_S$ obtained by partial tracing a random graph state \eqref{eq:parttrace} over the  subsystems labeled by  $T$. Here $\Ptrace = \{S,T\}$ denotes  a partition of the total number $n=2m$ of subspaces. For each block $C_i$ of $\Pvertex$, we define the sets $S_i = S \cap C_i$ and  $T_i = T \cap C_i$. These sets define a partition of each vertex $V_i$ and contain the information on which subsystems of the vertex $V_i$ are being traced out. Let $E_{i \to i} = \{ k \in C_i \; | \; \exists l \in C_i, \, l >k \text{ s.t. } (k,l) \in E\}$ be the set of bonds with both ends in $C_i$. Additionally, we have to take into account the bounds between the different blocks. For $j \neq i$ , let $E_{i \to j} = \{ k \in C_i \; | \; \exists l \in C_j \text{ s.t. } (i,j) \in E \}$. Notice that $|E_{i \to j}| = |E_{j \to i}|$ and that $d_{E_{i \to j}} = d_{E_{j \to i}}$. Since the edges of the graph are not oriented we shall put, for $i < j$, $E_{ij} = E_{i \to j}$.  Each unitary block $C_i$ is partitioned in two ways: by the partial tracing operation, $C_i = S_i \sqcup T_i$, and by the type of bonds it contains: $C_i = \sqcup_j E_{i \to j}$. 

\begin{figure}[htbp]
\centering
\subfigure[]{\label{fig:graph-example-marginal}\includegraphics{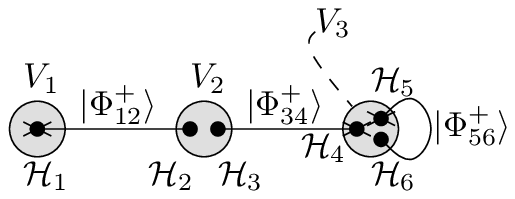}}\qquad\qquad
\subfigure[]{\label{fig:graph-example-gc}\includegraphics{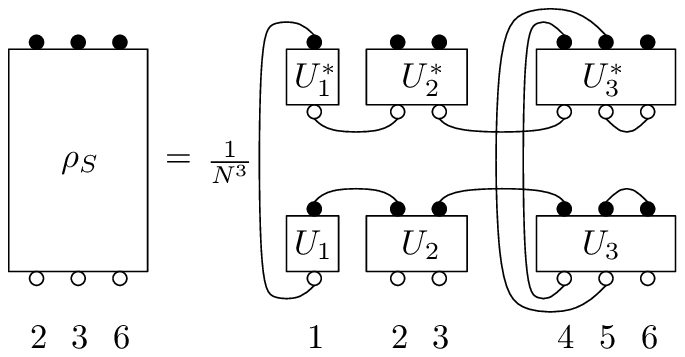}}
\caption{A marginal $\rho_S$ of the graph state represented in Figure \ref{fig:graph-example}, where Hilbert spaces $\H_1, \H_4$ and $\H_5$ have been traced out (a). The same marginal is represented using the graphical notation of \cite{collins-nechita-1} in (b).}

\end{figure}

For the case depicted in Figure \ref{fig:graph-example}, the vertex partition $\Pvertex$ is made of 3 blocks: $C_1 = \{1\}$, $C_2=\{2,3\}$ and $C_3 = \{4,5,6\}$. We represent it in Figure \ref{fig:graph-example-marginal}. For the block $C_3$ for example, one has $S_3 = \{6\}$, $T_3 = \{4,5\}$, $E_{3 \to 1}= \emptyset$, $E_{3 \to 2}= \{4\}$ and $E_{3 \to 3}= \{5,6\}$.

The marginal we are considering in Figure \ref{fig:graph-example-marginal} is $\rho_S = \trace_{\H_1 \otimes \H_4 \otimes \H_5}\ketbra{\Psi}{\Psi}$. In the diagrammatic picture of \cite{collins-nechita-1}, the density matrix $\rho_S$ is represented in Figure \ref{fig:graph-example-gc}.

In the general case, the constant in front of the diagram is meant to normalize the Bell states appearing in $\Psi$ and it has a value of $\sqrt{d_{[n]}N^n}$. Using the graphical Weingarten formula (Theorem \ref{thm:Wg_diag}), we obtain the following theorem, which contains the main result of this section. 

\begin{theorem}\label{thm:moments}
The moments of a graph state marginal $\rho_S$ are given by the exact formula
\begin{equation}\label{eq:moment_general_Wg}
\begin{split}
\E \trace(\rho_S^p) &= \left(d_{[n]}N^n\right)^{-p/2} \sum_{\substack{\alpha_1, \ldots, \alpha_k, \\ \beta_1, \ldots, \beta_k \in S_p}} \prod_{i=1}^k \left( d_{S_i}N^{|S_i|} \right)^{\#(\gamma^{-1} \alpha_i)} \prod_{i=1}^k \left( d_{T_i}N^{|T_i|} \right)^{\#\alpha_i} \\
&\cdot \prod_{1 \leq i < j \leq k} \left( d_{E_{ij}}N^{|E_{ij}|} \right)^{\#(\beta_i^{-1} \beta_j)} \prod_{i=1}^k \left( d_{E_{ii}}N^{|E_{ii}|} \right)^p \prod_{i=1}^k \Wg(d_{C_i}N^{|C_i|}, \alpha_i^{-1}\beta_i).
\end{split}
\end{equation}
\end{theorem}

\begin{proof}
For each independent unitary matrix $U_i$, $\left( d_{T_i}N^{|T_i|} \right)^{\#\alpha_i}$ counts the contribution of the loops corresponding to partial traces, $\left( d_{S_i}N^{|S_i|} \right)^{\#(\gamma^{-1} \alpha_i)}$ corresponds to the moment product, $\left( d_{E_{ii}}N^{|E_{ii}|} \right)^p$ corresponds to loops created by Bell states inside $V_i$ and, for each couple $i<j$, $\left( d_{E_{ij}}N^{|E_{ij}|} \right)^{\#(\beta_i^{-1} \beta_j)}$ represents the contribution of the loops coming from the Bell states between vertices $V_i$ and $V_j$. 
\end{proof}

The Weingarten functions appearing in equation \eqref{eq:moment_general_Wg} are intractable at fixed dimension $N$. In the following section, we shall investigate the marginals $\rho_S$ in the asymptotic regime $N \to \iy$ (keeping $d_1, d_2, \ldots, d_n$ fixed). 

In order to compute the dominating term in equation \eqref{eq:moment_general_Wg}, we look at the exponent of $N$ for each term in the sum (we use the first order asymptotic for the Weingarten function \eqref{eq:asympt-Wg} and the fact that for all permutations $\sigma \in \S_p$, $\#\sigma + |\sigma| = p$):
\begin{equation}
	-\frac{np}{2} + \sum_{i=1}^k \left[|S_i|(p-|\gamma^{-1}\alpha_i|) + |T_i|(p-|\alpha_i|)\right] + \sum_{1\leq i < j \leq k}|E_{ij}|(p-|\beta_i^{-1} \beta_j|) + \sum_{i=1}^k p|E_{ii}| + \sum_{i=1}^k |C_i|(-p-|\alpha_i^{-1}\beta_i|).
\end{equation}
Using, for all $i$, $|S_i| + |T_i| = |C_i|$, $\sum_{j\neq i} |E_{ij}| + 2|E_{ii}| = |C_i|$ and the fact that $\sum_i |C_i| = n$, we conclude that the general term in equation \eqref{eq:moment_general_Wg} has the following asymptotic behavior:
\begin{equation}
	N^{-F_{\alpha, \beta}} \left( \prod_{i=1}^k \Mob(\alpha_i^{-1} \beta_i) + o(1) \right),
\end{equation}
where 
\begin{equation}\label{eq:F-alpha-beta}
F_{\alpha, \beta} = \sum_{i=1}^k |S_i||\gamma^{-1}\alpha_i| + |T_i||\alpha_i| + \sum_{1\leq i < j \leq k} |E_{ij}||\beta_i^{-1} \beta_j| + \sum_{i=1}^k |C_i| |\alpha_i^{-1}\beta_i|.
\end{equation}
One has to minimize the function $F_{\alpha, \beta}$ over $\alpha_1, \ldots, \alpha_k, \beta_1, \ldots, \beta_k \in S_p$ in order to find the dominating term in equation \eqref{eq:moment_general_Wg}.
The following simplification can be made at this point:
\begin{lemma}
The minimum of the function $F_{\alpha, \beta}$ is the same as the minimum of the function $F_{\beta}$ defined by:
\begin{equation}\label{eq:F-beta}
 F_{\beta} = \sum_{i=1}^k |S_i||\gamma^{-1}\beta_i| + |T_i||\beta_i| + \sum_{1\leq i < j \leq k} |E_{ij}||\beta_i^{-1} \beta_j|.
\end{equation}
\end{lemma}

\begin{proof}
Since $|C_i| = |S_i| + |T_i|$, one may use $|S_i|$ times the triangular inequality $|\gamma^{-1}\alpha_i| + |\alpha_i^{-1}\beta_i| \geq |\gamma^{-1}\beta_i|$ and then $|T_i|$ times he triangular inequality $|\alpha_i| + |\alpha_i^{-1}\beta_i| \geq |\beta_i|$ to reduce the 
minimization problem of 
\eqref{eq:F-alpha-beta} to that of  \eqref{eq:F-beta}. The two problems have the same solution, since $F_\beta(\beta_1, \ldots, \beta_k) \leq F_{\alpha, \beta}(\alpha_1, \ldots, \alpha_k, \beta_1, \ldots, \beta_k)$ for all $\alpha_i$ and, choosing $\alpha_i = \beta_i$ for all $i$, one has $F_\beta = F_{\alpha, \beta}$.
\end{proof}

This time, $F_\beta$ is a function which depends only on $k$ permutations $\beta_1, \ldots, \beta_k \in \S_p$, hence minimizing $F_\beta$ should be easier than minimizing $F_{\alpha,\beta}$. In the next section, we will present a complete solution to the minimization problems \eqref{eq:F-alpha-beta} and \eqref{eq:F-beta} using a graph-theoretical method.

\subsection{The minimization problem on permutations as a maximal flow problem}
\label{sec:flow}

We will now exhibit a connection between the minimization problems \eqref{eq:F-alpha-beta} 
and \eqref{eq:F-beta} and a maximum flow problem on a network. For an introduction to the maximum flow on networks, 
see \cite{cormen}, Chapter 27. We introduce a network  $(\mathcal V, \mathcal E, w)$ with vertex set $\mathcal V$,
 edge set $\mathcal E$ and edge capacities $w$. The network is associated to the minimization problem for $F_\beta$ \eqref{eq:F-beta} 
in the following way. The vertex set is given by $\mathcal V  = \{\id, \gamma, \beta_1, \ldots, \beta_k\}$, with two distinguished vertices: 
the \emph{source} $s = \id$ and the \emph{sink} $t=\gamma$.  The edges in $\mathcal E$ are oriented and they are of three types:
\begin{equation}
\mathcal E = \{(\id, \beta_i) \; ; \; |T_i| >0\} \sqcup \{(\beta_i, \gamma) \; ; \; |S_i| >0\} \sqcup \{(\beta_i, \beta_j), (\beta_j, \beta_i) \; ; \; |E_{ij}| >0\}.
\end{equation}
The capacities of the edges are given by:
\begin{equation}
\begin{split}
&w(\id, \beta_i) = |T_i| > 0\\
&w(\beta_i, \gamma) = |S_i| > 0\\
&w(\beta_i, \beta_j) = w(\beta_j, \beta_i) = |E_{ij}| >0.
\end{split}
\end{equation}

\begin{figure}[htbp]
\centering
\includegraphics{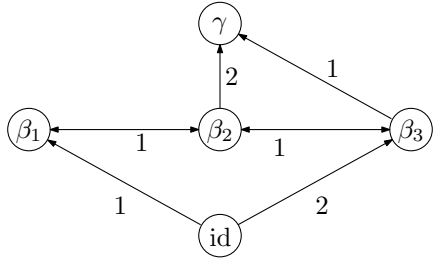}
\caption{Network corresponding to the graph state marginal of Figure \ref{fig:graph-example-marginal}.}
\label{fig:graph-example-network}
\end{figure}

The network corresponding to Figures \ref{fig:graph-example-marginal} and \ref{fig:graph-example-gc} 
is represented in Figure \ref{fig:graph-example-network}.

Let us now consider the maximum flow problem on the network $(\mathcal V, \mathcal E, w)$, with source $s=\id$ and sink $t=\gamma$. A \emph{flow} in a network is a function $f:\mathcal V \times \mathcal V \to \R$ with the following three properties:
\begin{enumerate}
\item \textbf{Capacity constraint}: for all vertices $u,v \in \mathcal V$, $f(u,v) \leq w(u,v)$;
\item \textbf{Skew symmetry}: for all  $u,v \in \mathcal V$, $f(u,v) = -f(v,u)$;
\item \textbf{Flow conservation}: for all vertices $u$ different from the source and the sink $u \in \mathcal V \setminus \{s,t\}$, 
\begin{equation}
\sum_{v \in \mathcal V} f(u,v) = 0.
\end{equation}
\end{enumerate}

The value of a flow $f$ is the quantity
\begin{equation}
|f| = \sum_{u \in \mathcal V} f(s,v).
\end{equation}

In the maximum flow problem, we want to determine a flow of maximum value in the network $(\mathcal V, \mathcal E, w)$. The Ford-Fulkerson algorithm states that solutions to the maximum flow problem are obtained as sums of \emph{augmenting paths}. One starts with an empty flow $f \equiv 0$. An augmenting path is obtained by pushing a quantity $x$ of flow through a path $\id \to \beta_{i_1} \to \beta_{i_2} \to \cdots \to \beta_{i_l} \to \gamma$, such that, for all edges $(u,v)$ of the path, $x \leq w(u,v)$. One can then update the flow function by adding $x$ units of flow to each of $f(\id, \beta_{i_1}), f(\beta_{i_1}, \beta_{i_2}), \ldots, (\beta_{i_l}, \gamma)$. The weights are also updated by subtracting $x$ from the capacities of the edges above. The new network, with updated capacities, is called the \emph{residual network}. One iterates this procedure until no more augmenting paths exists. The maximal flow that can be sent from $s$ to $t$ is denoted by $X = |f|$. The final residual network is denoted by $(\mathcal V_\text{res}, \mathcal E_\text{res}, w_\text{res})$. Note that $\mathcal E_\text{res}$ contains only edges for which $w_\text{res} >0$. 

Let us analyze now the connection between \emph{a solution} of the maximal flow problem and $F_\beta$. Each time a flow $x$ is sent through a path $\id \to \beta_{i_1} \to \beta_{i_2} \to \cdots \to \beta_{i_l} \to \gamma$, we write the triangular inequality
\begin{equation}\label{eq:network-triangular-inequality}
x\left[ |\beta_{i_1}| + |\beta_{i_1}^{-1} \beta_{i_2}| + \cdots + |\beta_{i_{l-1}}^{-1} \beta_{i_l}| + |\beta_{i_l}^{-1} \gamma| \right] \geq x |\gamma| = x(p-1).
\end{equation}
Summing up all this inequalities for the augmenting paths that add up to the maximal flow $X$, we have that
\begin{equation}\label{eq:residual-network}
\begin{split}
F_\beta &\geq X(p-1) + \sum_{(\id, \beta_i) \in \mathcal E_\text{res}} w(\id, \beta_i)|\beta_i| + \sum_{(\beta_i, \gamma) \in \mathcal E_\text{res}} w(\beta_i, \gamma)|\beta_i^{-1} \gamma| \\
&+\sum_{(\beta_i, \beta_j) \in \mathcal E_\text{res},\;  i<j} w(\beta_i, \beta_j)|\beta_i^{-1}\beta_j|.
\end{split}
\end{equation}

Moreover, we claim that equality can be achieved in $F_\beta \geq X(p-1)$.
 The residual network has the property that it contains no augmenting paths.
 This means that the source $\id$ and the sink $\gamma$ are in different connected components of the graph $(\mathcal V_\text{res}, \mathcal E_\text{res})$. The equality case in $F_\beta \geq X(p-1)$ is obtained by setting $\beta_i = \id$ for all permutations $\beta_i$ in the connected component of the source $\id$. Similarly, put $\beta_i = \gamma$ for all vertices $\beta_i$ in the connected component of the sink $\gamma$. Finally, impose that $\beta_i = \beta_j$ for all permutations $\beta_i, \beta_j$ lying in a connected component different from the ones containing $\id$ of $\gamma$. In this manner, all extra terms in equation \eqref{eq:residual-network} are zero, and one has indeed $F_\beta = X(p-1)$. 

In conclusion, we have showed that $F_\beta \geq X(p-1)$ always holds, where $X$ is the maximum flow on the network $(\mathcal V, \mathcal E, w)$. Of course, $X$ does not depend to the solution $\mathcal F$ (i.e. the choice of the augmenting paths leading to the maximum flow) to the maximum flow problem. Moreover, we have showed that for each solution $\mathcal F$, equality can be attained by choosing permutations $\beta_i$ in a way depending on the residual network (which depends on the solution $\mathcal F$). Let us now describe exactly, for a solution $\mathcal F$ to the maximum flow problem, the set of $k$-tuples $(\beta_1, \ldots, \beta_k) \in \S_p^k$ which saturate the inequality $F_\beta \geq X(p-1)$. 

Firstly, each time an augmenting path $\id \to \beta_{i_1} \to \beta_{i_2} \to \cdots \to \beta_{i_l} \to \gamma$ in $\mathcal F$ is chosen, we use the triangular inequality \eqref{eq:network-triangular-inequality}. This implies that \emph{all} permutations $\beta_{i_1}, \beta_{i_2}, \ldots, \beta_{i_l}$ are geodesic permutations (i.e. elements of $\S_{NC}(p)$ and that they satisfy
\begin{equation}\label{eq:linear-ordering-geodesic-permutations}
	[\beta_{i_1}] \leq [\beta_{i_2}] \leq \cdots \leq [\beta_{i_l}],
\end{equation}
where we note by $[\sigma]$ the cycle partition of a permutation. Actually, since we are dealing with geodesic permutations, 
it follows 
(see e.g. the last paragraph of section \ref{sec:notations}), 
that these partitions are non-crossing. Hence, each augmenting path imposes two conditions on the permutations it contains: they should all be geodesic permutations and their associated  non-crossing partitions should satisfy a linear ordering inequality \eqref{eq:linear-ordering-geodesic-permutations}. Moreover, the condition that the extra terms in equation \eqref{eq:residual-network} should be zero imposes additional constraints on the permutations (except, of course, in the case where $\mathcal E_\text{res}$ is empty, and all connected components of the residual network are singletons). These conditions impose that all permutation lying in the same connected component of 
$(\mathcal V_\text{res}, \mathcal E_\text{res})$ should be equal (so that the terms $|\beta_i^{-1}\beta_j|$ are all null). Hence, the permutations $\beta_i$ achieving the minimum for $F_\beta$, for a solution $\mathcal F$ of the maximum flow problem, should satisfy the following constraints (note that we are referring to the connected components of the final residual network $(\mathcal V_\text{res}, \mathcal E_\text{res})$):
\begin{equation}\label{eq:constraints-permutations}
\begin{split}
[\beta_{i_1}] \leq [\beta_{i_2}] \leq \cdots \leq [\beta_{i_l}],\qquad &\text{ for all augmenting paths in the solution $\mathcal F$;}\\
\beta_i = \id, \qquad &\text{ for all vertices $\beta_i$ in the connected component of $\id$;}\\
\beta_i = \gamma, \qquad &\text{ for all vertices $\beta_i$ in the connected component of $\gamma$;}\\
\beta_i = \beta_j, \qquad &\text{ for all vertices $\beta_i, \beta_j$ in the same connected component.}\\
\end{split}
\end{equation}

Two important comments should be made at this point. Using the properties of the augmenting flow, it is fairly easy to see that the above system of equations and inequalities is consistent. Moreover, all solutions of \eqref{eq:constraints-permutations} are \emph{geodesic} permutations. We have seen that this already holds for permutations $\beta_i$ which belong to at least one augmenting path. A permutation not belonging to any augmenting paths is connected in the residual network to either the source $\id$ or to the sink $\gamma$ (but not to both, since in that case $\id \to \beta_i \to \gamma$ will be a valid augmenting path in the final residual network, which is impossible). In that case, $\beta_i = \id$ or $\beta_i = \gamma$, which are both geodesic permutations. 

Let $\tilde B_{\mathcal F}$ be the set of solutions to \eqref{eq:constraints-permutations}. It can be described by a poset $\mathcal P_{\mathcal F}$, endowed with a partial order $\tilde\prec_{\mathcal F}$:
\begin{equation}
\tilde B_{\mathcal F} = \{ \beta_1, \ldots, \beta_k \in \S_{NC}(p)^k \; | \; [\beta_i] \leq [\beta_j] \text{ whenever } i \tilde\prec_{\mathcal F} j \}.
\end{equation}

Since there are more than one possible solutions to the maximum flow problem (all having the same value $X$), the general solution to the minimization problem \eqref{eq:F-beta} is given by the union over possibilities:
\begin{equation}\label{eq:final-solution-beta}
\tilde B = \bigcup_{\substack{\mathcal F \text{ solution to the}\\ \text{Max-Flow problem}}} \tilde B_{\mathcal F}.
\end{equation}

The solution of the minimization problem for the $\beta$'s being settled, let us move now to the initial minimization problem for $F_{\alpha,\beta}$ over $\alpha_1, \ldots, \alpha_k, \beta_1, \ldots, \beta_k$, stated in equation \eqref{eq:F-alpha-beta}. We solve the problem completely, by characterizing all the permutations $\alpha_1, \ldots, \alpha_k$ which achieve the minimum. We shall consider three cases for a vertex $i$ (which depend only on the initial network):
\begin{enumerate}
\item[(I)] $|S_i| >0, |T_i|>0$: the vertex $\beta_i$ is connected to both the source and the sink in the network $(\mathcal V, \mathcal E, w)$;
\item[(II)] $|S_i| = 0, |T_i|>0$: the vertex $\beta_i$ is connected only to the source in the network $(\mathcal V, \mathcal E, w)$;
\item[(III)] $|T_i| = 0, |S_i|>0$: the vertex $\beta_i$ is connected only to the sink in the network $(\mathcal V, \mathcal E, w)$.
\end{enumerate}
\begin{definition}
A vertex $V_i$ is said to be of \emph{type ``T''} if all the subsystems of $V_i$ are traced out. A vertex $V_j$ is called a \emph{type ``S'' vertex} if none of the subsystems of $V_j$ are traced out.
\end{definition}

In the first case, when going from the $F_{\alpha, \beta}$-minimization problem to the $F_\beta$-minimization problem, we used \emph{both} triangle inequalities $|\gamma^{-1}\alpha_i| + |\alpha_i^{-1}\beta_i| \geq |\gamma^{-1}\beta_i|$ and  $|\alpha_i| + |\alpha_i^{-1}\beta_i| \geq |\beta_i|$ at least once. For a fixed vertex $i$ and a fixed permutation $\beta_i$, the unique solution to the system
\begin{equation}\label{eq:alpha_i-beta_i}
\begin{split}
&|\gamma^{-1}\alpha_i| + |\alpha_i^{-1}\beta_i| = |\gamma^{-1}\beta_i| \\
&|\alpha_i| + |\alpha_i^{-1}\beta_i| = |\beta_i|
\end{split}
\end{equation}
is $\alpha_i = \beta_i$. This follows from the following chain of inequalities:
\begin{equation}\label{eq:alpha_i-beta_i-chain}
p-1 + 2|\alpha_i^{-1}\beta_i| \leq |\gamma^{-1}\alpha_i| + |\alpha_i^{-1}\beta_i| + |\alpha_i| + |\alpha_i^{-1}\beta_i| = |\gamma^{-1}\beta_i| + |\beta_i| = p-1,
\end{equation}
and hence $|\alpha_i^{-1}\beta_i| = 0$, for all vertices of type (I).

In the case (II) above, since $|S_i| = 0$, only the triangle inequality 
$|\alpha_i| + |\alpha_i^{-1}\beta_i| = |\beta_i|$ must be saturated. Hence, for a given geodesic permutation $\beta_i$, the set of permutations $\alpha_i$ which saturate the triangle inequality is the geodesic set $\id \to \alpha_i \to \beta_i$. In other words, we have to consider all the geodesic permutations $\alpha_i \in \S_{NC}(p)$ such that $[\alpha_i] \leq [\beta_i]$. By considering all the possibilities for $\alpha_i$, one needs to compute the sum
\begin{equation}
\sum_{[\alpha_i] \leq [\beta_i]} \Mob(\alpha_i^{-1}\beta_i).
\end{equation}
Using the fact that the function $\Mob$ is related to the M\"{o}bius function on the poset of non-crossing partitions, one can show (see \cite{nica-speicher}, chapter 10 and equation \eqref{eq:inversion-Mob}) that the above sum is non-zero only if the above sum is trivial, that is, only if $\beta_i = \id$. Thus, permutations $\alpha_i, \beta_i$ corresponding to a vertex which is connected only to the source (i.e. a type ``T'' vertex) \emph{must} satisfy $\alpha_i = \beta_i = \id$. Similar ideas lead to the conclusion that all type ``S'' vertices (case (III) above) need to satisfy $\alpha_i = \beta_i = \gamma$. Moreover, we have shown, in all three cases (I)-(III) above, that the permutations $\alpha_1, \ldots, \alpha_k$ which achieve the minimum satisfy $\alpha_i = \beta_i$, for all $i=1, \ldots, k$. This greatly simplifies equation \eqref{eq:F-alpha-beta}, since the M\"{o}bius functions are trivial: $\Mob(\alpha_i^{-1}\beta_i) = \Mob(\id) = 1$.

In conclusion, since some of the solutions in equation \eqref{eq:final-solution-beta} cancel out, we define, for each solution $\mathcal F$ of the maximum flow problem, the set
\begin{equation}
B_{\mathcal F} = \{ \beta_1, \ldots, \beta_k \in \tilde B_{\mathcal F}\; | \; \beta_i=\id \text{ for type ``T'' vertices and } \beta_j=\gamma \text{ for type ``S'' vertices} \}.
\end{equation}
We introduce the modified (smaller) set of  general solutions to the minimization problem:
\begin{equation}
B = \bigcup_{\substack{\mathcal F \text{ solution to the}\\ \text{Max-Flow problem}}} B_{\mathcal F}.
\end{equation}

We sum up the preceding discussion in the following theorem, the main result of this section. 

\begin{theorem}\label{thm:moments-network}
\begin{equation}
\E \trace(\rho_S^p) = N^{-X(p-1)} \sum_{(\beta_1, \ldots, \beta_k) \in B} \prod_{i=1}^k \left( d_{S_i} \right)^{\#(\gamma^{-1} \beta_i)} \prod_{i=1}^k \left( d_{T_i} \right)^{\#\beta_i}
\cdot \prod_{1 \leq i < j \leq k} \left( d_{E_{ij}} \right)^{\#(\beta_i^{-1} \beta_j)} \prod_{i=1}^k d_{C_i}^{-p}.
\end{equation}
\end{theorem}

In the simplified situation when the parameters are trivial, $d_i = 1 \; \forall i$, the above theorem admits the following corollary, which provides a simple combinatorial formula for the moments of a graph state marginal. Both combinatorial quantities below (the maximum flow $X$ and the cardinality of the set of solutions $\# B$) can be computed from the network associated to the marginal.

\begin{theorem}\label{thm:moments-network-N}
Consider a graph state $\ket{\Psi}$ with the property that the relative dimensions of its subsystems are unity ($d_1 = \cdots = d_n = 1$). The average moments of a marginal $\rho_S = \trace_T \ketbra{\Psi}{\Psi}$ are given by the simple combinatorial formula
\begin{equation}
	\E \trace(\rho_S^p) = N^{-X(p-1)} (\# B + o(1)),
\end{equation}
where $X$ is the maximum flow associated to the marginal and $B$ is determined by the set of augmenting paths corresponding to the maximum flow problem.
\end{theorem}

Theorem \ref{thm:moments-network-N} is very convenient in the sense that it turns the problem of computing moments into the problem of counting the number of solutions of a maximum flow problem. 
This is a rather unexpected mathematical connection between quantum physics and networking theory.

\begin{remark}
When one interchanges the elements of the partition defining the
 marginal $S \leftrightarrow T$, all the objects in the above discussion are replaced by their duals. 
In the network, the source and the sink are exchanged; the value of the maximum flow does not change, 
and all the inequalities for the $\beta$'s are reversed. This means that geodesic permutations $\alpha_i$ and
 $\beta_i$ are replaced by their Kreweras complement. Hence, the asymptotic moments do not change. This was expected, since the spectra of $\trace_I \ketbra{\Psi}{\Psi}$ and $\trace_{I^c} \ketbra{\Psi}{\Psi}$ differ only by null eigenvalues.
\end{remark}

In this paper, we are making heavy use of the Fubin-Study measure. It is the most natural one, as it ensures norm preservation of the
states. However it is not the simplest one, and it would be computationally simpler to deal with random Ginibre ensembles instead
of unitary matrices. The following observation shows that the results are asymptotically the same:

\begin{remark}
The graphical calculus that we recalled in section \ref{sec:planar} has a Gaussian counterpart \cite{collins-nechita-3}, and it follows by rather direct 
inspection that the Theorems \ref{thm:moments-network} and \ref{thm:moments-network-N} would yield the same asymptotic
results if one replaced in our model unitary matrices by (properly normalized) non-hermitian standard complex Gaussian random matrices. 
Non-hermitian standard complex Gaussian matrices are less natural (and in particular they are not norm preserving). However
the graphical calculus is much simpler, as the Weingarten calculus has just to be replaced by Wick theorem (i.e. Feynman diagrams).
The reason why one obtains the same result, is that for the Gaussian case, one does not have $\alpha$'s and $\beta$'s but 
just $\beta$'s and one ends up directly with the minimization problem for $F(\beta )$.
The fact that the leading coefficients are the same follows from the non-crossing M\"{o}bius inversion formula \eqref{eq:inversion-Mob}. 
\end{remark}
We are not able to give physical reasons why the Gaussian and the unitary model have the same asymptotic behavior in general and it seems non-obvious to us without a direct computation of moments.

\section{Applications}\label{sec:applications}

In the last part of the paper, we study applications of the theory developed in the preceding sections. We start by introducing the Fuss-Catalan probability distributions which appear as limit eigenvalue distributions in some special cases. We then study particular graphs (stars graphs, cycles) and the density matrix ensembles one obtains by partial tracing the corresponding random pure states. Finally, in Section \ref{sec:exotic}, we study marginals of graph-states which lead to new probability distributions.

\subsection{Free Poisson and Fuss-Catalan distributions}

In Random Matrix Theory, the \emph{free Poisson} (or \emph{Marchenko-Pastur}) distribution $\pi^{(1)}_c$ describes asymptotically the spectral density of the normalized random Wishart matrices $W=GG^{*} \in \M_N(\C)$, where $G$ is a random rectangular matrix from the Ginibre ensemble of size $N \times cN$. The Marchenko-Pastur distribution is also the limit spectral density of rescaled density matrices from the \emph{induced ensemble} \cite{nechita, BZ06}.
If $\lambda$ denotes an eigenvalue of a normalized random Wishart density matrix $W/\trace W$ of size $N$
the probability density of the variable $x=N\lambda$ is 
asymptotically given by the 
free Poisson probability measure with parameter $c>0$,
\begin{equation}
\label{eq:freepoiss}
\pi^{(1)}_c=\max (1-c,0)\delta_0+\frac{\sqrt{4c-(x-1-c)^2}}{2\pi x} \; \mathbf{1}_{[1+c-2\sqrt{c},1+c+2\sqrt{c}]}(x) \; dx \; ,
\end{equation}
where $\mathbf{1}_A$ denotes the indicator function of the set $A$.

In free probability theory, this distribution is also called the \emph{free $\chi^2$ distribution}, and it has a semigroup structure with respect to the additive free convolution of Voiculescu: $\pi^{(1)}_c \boxplus \pi^{(1)}_d = \pi^{(1)}_{c+d}$ (see, e.g. \cite{nica-speicher}). Moreover, $\pi^{(1)}_c$ can be characterized by the fact that all its free cumulants are equal to $c$.

The mean purity for the Marchenko-Pastur distribution reads $\int x^2 \; d\pi^{(1)}_c(x) = c^2+c$.
One can compute the mean entropy of this probability distribution (note that in this work, logarithms are considered in base 2, $\log = \log_2$)
\begin{equation}
\label{eq:entropy-free-poisson}
H(\pi^{(1)}_c) \:= \  \int_{}^{} -x \log x \; d\pi^{(1)}_c(x) = 
\begin{cases}
-\frac{1}{2} - c \log c \quad & \text{ if } c \geq 1,\\
-\frac{c^2}{2} \quad & \text{ if } 0<c<1.
\end{cases}
\end{equation}
In the case $c=1$ it is equal to $-1/2$, so
the mean entropy of a random density matrix $\rho$ 
of size $N$ generated out of a square Ginibre matrix $G \in \M_N(\C)$ behaves asymptotically ($N \to \iy$) as $\log N- 1/2$.

Next, we recall a few facts about a series of probability distributions that was discovered recently in relation to quantum group theory in \cite{banica-etal}, and which generalize the free Poisson distribution \eqref{eq:freepoiss}. These distributions depend on a real parameter $s>0$ and are most easily characterized by their moments. Four our purposes it is convenient to extend the meaning of the binomial notation: for an arbitrary real number $\alpha$
and a natural integer $k$ we set 
\begin{equation}
\binom{\alpha}{k}
 =  \frac{\alpha(\alpha-1)\ldots(\alpha-k+1)}{k!} .
\end{equation}

\begin{theorem}
For any real number $s>0$, there exists a probability measure $\pi^{(s)}$, called the Fuss-Catalan distribution of order $s$, whose moments are the generalized
Fuss-Catalan  numbers
(see, e.g. \cite{armstrong}):
\begin{equation}
\label{eq:FCmoments}
FC^{(s)}_p=\frac{1}{sp+1}\binom{sp+p}{p}  .
\end{equation}

The measure $\pi^{(s)}$ has no atoms, it is supported on $[0,K]$ where $K=(s+1)^{s+1}/s^s$, its density is analytic on $(0,K)$, and bounded at $x=K$, with asymptotic behavior $\sim 1/(\pi x^{s/(s+1)})$ at $x=0$.
\end{theorem}

This distribution arises in Random Matrix Theory when one studies the product of $s$ independent random square Ginibre matrices, $G=\prod_{j=1}^s G_j$. In this case, $s \in \N$ and squared singular values of $G$ (i.e. eigenvalues $GG^*$) have asymptotic distribution $\pi^{(s)}$. In terms of free probability theory, it is the free multiplicative convolution product of $s$ copies of a Marchenko-Pastur distribution \cite{banica-etal}:
\begin{equation}
\pi^{(s)} = \left(\pi^{(1)}\right)^{\boxtimes s}.
\end{equation}

One can further generalize Fuss-Catalan distributions in the following way. Consider $s$ independent rectangular Ginibre matrices $G_1, \ldots, G_s$ of dimension ratios $c_1, \ldots, c_s$ and such that the product $G=\prod_{j=1}^s G_j$ is well defined. The asymptotic level distribution for the (rescaled) eigenvalues of $GG^*$ is then $\pi^{(s)}_{\bf c}$. The new, generalized measure depends on $s$ parameters ${\mathbf c}=\{c_1,\dots c_s\}$
which describe the dimension ratios for each matrix. 
In free probability this corresponds to taking $s$ Marchenko-Pastur distributions
with different parameters $c$:
\begin{equation}
\pi^{(s)}_{\mathbf c} = \stackrel[{j=1}]{s}{\boxtimes} \pi^{(1)}_{c_j}.
\end{equation}

In the case of an integer $s$, the Fuss-Catalan numbers count the number of $s$-chains in the lattice of non-crossing partitions $NC(p)$ \cite{armstrong}:
\begin{equation}\label{eq:def-comb-FC}
FC^{(s)}_p = | \{\hat 0_p \leq \sigma_1 \leq \sigma_2 \leq \cdots \leq \sigma_s \leq \hat 1_p\in NC(p)\}|.
\end{equation}
In \cite{banica-etal}, the authors interpreted these numbers as the moments of \emph{free Bessel laws}. They showed that these numbers count non-crossing partitions of $[sp]$ into blocks of size multiple of $s$ (for a bijective proof of this statement, see \cite{armstrong}):
\begin{equation}
FC^{(s)}_p = |\{ \tau \in NC(sp) \; | \; \forall b \in \tau, \; 
|b|\text{ is a multiple of }s\}|.
\end{equation}

Using the moment formulas, one can compute the entropy of the Fuss-Catalan distributions. 

\begin{proposition}
\label{prop:entropy-Fuss-Catalan}
The entropy $H$ of the Fuss-Catalan measure $\pi^{(s)}$ is given by
\begin{equation}
H(\pi^{(s)}) \ := \ 
\int -x \log x \; d\pi^{(s)}(x) = - \sum_{j=2}^{s+1} \frac{1}{j}.
\end{equation}
\end{proposition}
\begin{proof}
Using the analyticity of the density, we can write the Shannon entropy as a limit of R\'enyi entropies,
 by replacing factorials in the definition of $FC^{(s)}_p$ by Gamma functions:
\begin{equation}
\int -x \log x \; d\pi^{(s)}(x) = \lim_{p \downarrow 1} \frac{1}{1-p} \log \frac{\Gamma(sp+p+1)}
{\Gamma(sp+2) \Gamma(p+1)}.
\end{equation}
Next, we write the above limit as a logarithmic derivative at $p=1$ of the function
\begin{equation}
p \mapsto \frac{\Gamma(sp+p+1)}{\Gamma(sp+2) \Gamma(p+1)}.
\end{equation}
Using the expression of the derivative of Gamma function at integer points
\begin{equation}\Gamma'(n+1) = n! \left( -\gamma_0 + \sum_{j=1}^n \frac{1}{j} \right),\end{equation}
where $\gamma_0 \approx 0.57721$ is the Euler constant, we can conclude.
\end{proof}

The second member of the Fuss-Catalan family, $\pi^{(2)}$ corresponds to a product of $s=2$
 independent Ginibre matrices. Note that a 
related analysis of a product of two real random correlation matrices
was performed by  Bouchaud et al. \cite{BLMM07}, who used the term 
\emph{Marchenko-Pastur square distribution},
and independently by Benaych-Georges \cite{BG09},
who studied multiplicative convolution of  free Poisson laws.
The formulas in equation \eqref{eq:FCmoments} for $s=2$ allow to explicitly solve for the density of this measure:
\begin{equation}
d\pi^{(2)}(x) = \frac{\sqrt[3]{2} \sqrt{3}}{12 \pi} \;
 \frac{\sqrt[3]{2} \left(27 + 3\sqrt{81-12x} \right)^{\frac{2}{3}} - 6\sqrt[3]{x}}{x^{\frac{2}{3}}\left(27 + 3\sqrt{81-12x} \right)^{\frac{1}{3}}} 
  \; \mathbf{1}_{(0, \frac{27}{4}]}(x) \; dx.
\end{equation}

\begin{figure}[htbp]
\centering
\includegraphics[width=0.6\textwidth]{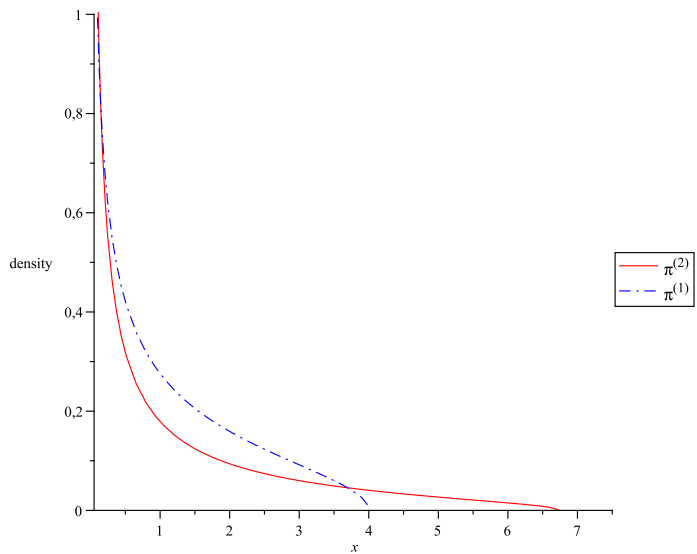}
\caption{Plot of the densities for the probability measures $\pi^{(2)}$ and $\pi^{(1)}$.}
\label{fig:plot-density}
\end{figure}

A plot of the density of $\pi^{(2)}$ is given in 
Figure \ref{fig:plot-density}, 
along with the density of the free Poisson distribution $\pi^{(1)}$. 

\subsection{Marginals of graphs leading to Fuss-Catalan distributions} 
We provide in this section examples of simple graphs and associated marginals whose limit distributions are members of the Fuss-Catalan family of probability laws introduced in the preceding section. The examples chosen here have the property of being the simplest ones which lead to Fuss-Catalan limits. 

\begin{figure}[htbp]
\centering
\subfigure[]{\includegraphics{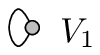}}\qquad\qquad
\subfigure[]{\includegraphics{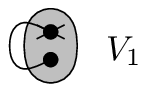}}\quad\quad
\subfigure[]{\includegraphics{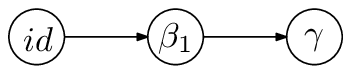}}
\caption{A vertex with one loop (a) and a marginal (b) having as a limit eigenvalue distribution the Marchenko-Pastur law $\pi^{(1)}$. In the network (c), both edges have capacity one.}
\label{fig:pi_1}
\end{figure}

The simplest case, $s=1$, corresponds to the free Poisson (or Marchenko-Pastur) distribution.
 This distribution is the limit of the \emph{induced ensemble} 
of random density matrices studied in \cite{BZ06, nechita}. 
The simplest graph having the Marchenko-Pastur distribution as a limit, is the one-loop graph, 
with one vertex containing two subsystems (or Hilbert spaces) inside.
 The marginal we consider is the one where we trace out one system, keeping the other 
(since we assume that all the Hilbert spaces are isomorphic to $\C^N$, the indices of the systems
 being traced out do not matter). The simple graph, the marginal and the associated network are
 represented in Figure \ref{fig:pi_1}. The maximum flow problem associated to the network is trivial, 
one unit of flow being sent through the path $\id \to \beta_1 \to \gamma$. The residual network is empty and it follows from Theorem \ref{thm:moments-network} that
\begin{equation}
\E \trace \rho_S^p \sim N^{1-p} \cdot |\{[\beta_1] \in NC(p) \; | \; \hat 0_p \leq [\beta_1] \leq \hat 1_p\}| \qquad \forall p \geq 1.
\end{equation}
The number of all non-crossing partitions of $[p]$ is known to be 
the $p$-th Catalan number $FC^{(1)}_p$, 
which is also the $p$-th moment of the Marchenko-Pastur distribution $\pi^{(1)}$.
By construction both subsystems are of the same dimension
so if one performs partial trace over one of them
the resulting  mixed state is distributed according to the
Hilbert-Schmidt measure \cite{SZ04,BZ06}
and the ratio parameterizing the Marchenko-Pastur distribution reads $c=1$.

\begin{figure}[htbp]
\centering
\begin{minipage}[c]{.46\linewidth}\centering
\subfigure[]{\includegraphics{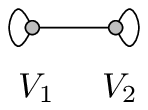}}\\
\subfigure[]{\includegraphics{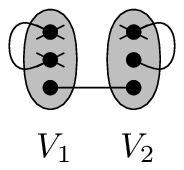}}
\end{minipage} \hfill
\begin{minipage}[c]{.46\linewidth}\centering
\subfigure[]{\includegraphics{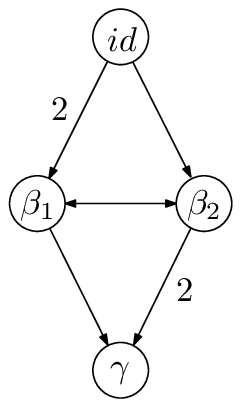}}
\end{minipage}
\caption{A graph (a) and a marginal (b) having as a limit eigenvalue distribution the Fuss-Catalan law $\pi^{(2)}$. In the network (c), non-labeled edges have capacity one.}
\label{fig:pi_2}
\end{figure}

We now move on to the first non-trivial case, $s=2$. The graph we consider (see Figure \ref{fig:pi_2}) has two vertices connected by an edge and each having a loop attached. The marginal of interest is the one obtained by partial tracing two copies of $\C^N$ in the first vertex $V_1$ and one copy of $\C^N$ in $V_2$. In the network associated to this marginal, a maximum flow of 3 can be sent from the source $\id$ to the sink $\gamma$: one unit through each path $\id \to \beta_i \to \gamma$, $i=1,2$ and one unit through the path $\id \to \beta_1 \to \beta_2 \to \gamma$. In this way, the residual network is empty and the only constraint on the geodesic permutations $\beta_1, \beta_2$ is 
\begin{equation}
\hat 0_p \leq [\beta_1] \leq [\beta_2] \leq \hat 1_p,
\end{equation}
i.e. $[\beta_1]$ and $[\beta_2]$ form a 2-chain in $NC(p)$. It follows the moments of the marginal $\rho_S$ are given by
\begin{equation}
\E \trace \rho_S^p \sim N^{3(1-p)} FC_p^{(2)},
\end{equation}
and thus the rescaled random matrix $N^3\rho_S$ converges in distribution to the second Fuss-Catalan measure $\pi^{(2)}$. 

\begin{figure}[htbp]
\centering
\begin{minipage}[c]{.46\linewidth}\centering
\subfigure[]{\label{fig:pi_s_general_simple}\includegraphics{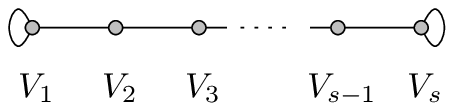}}\\
\subfigure[]{\label{fig:pi_s_general}\includegraphics{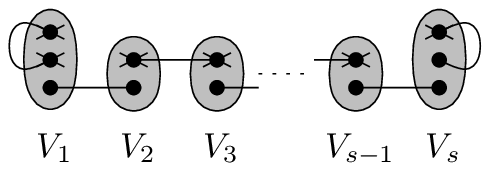}}
\end{minipage} \hfill
\begin{minipage}[c]{.46\linewidth}\centering
\subfigure[]{\label{fig:pi_s_general_network}\includegraphics{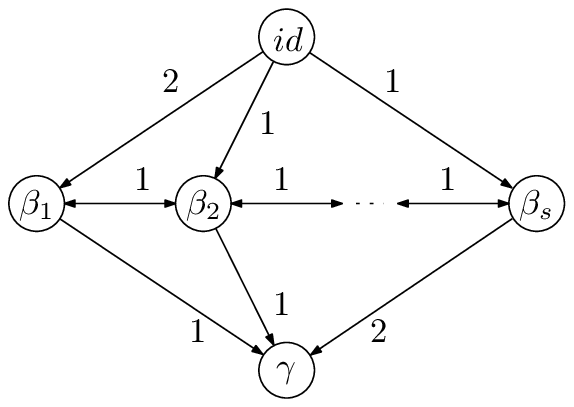}}
\end{minipage}
\caption{An example of a graph state (a) with a marginal (b) having as a limit eigenvalue distribution the $s$-th Fuss-Catalan probability measure $\pi^{(s)}$. The associated network (c) has a maximal flow of $s+1$, obtained by sending a unit of flow through each $\beta_i$ and a unit through the path $\id \to \beta_1 \to \cdots \to \beta_s \to \gamma$. The linear chain condition $[\beta_1] \leq \cdots \leq [\beta_s]$ follows.}
\label{fig:pi_s_general_example}
\end{figure}

The construction for the graph state and its marginal in Figure \ref{fig:pi_2} can be easily generalized to obtain graph state marginals having $\pi^{(s)}$ as a limit distribution. The main idea behind this construction is the structure of the network in Figure \ref{fig:pi_s_general_network}. The maximum flow in this network is $s+1$ and it can be obtain by sending one unit of flow through the following paths:
\begin{equation}
\begin{split}
&\id \to \beta_1 \to \gamma\\
&\id \to \beta_2 \to \gamma\\
&\cdots \\
&\id \to \beta_s \to \gamma\\
&\id \to \beta_1 \to \beta_2 \to \cdots \to \beta_s \to \gamma
\end{split}
\end{equation}

The residual network is empty, and thus the combinatorial factor appearing in equation \eqref{eq:moment_general_Wg} is given by the number of $s$-chains in $NC(p)$. In conclusion, the moments of $\rho_S$ behave asymptotically as
\begin{equation}
\E \trace \rho_S^p \sim N^{(s+1)(1-p)} FC_p^{(s)}.
\end{equation}
The empirical eigenvalue distribution of the rescaled marginal $N^{s+1} \rho_S$ converges to the Fuss-Catalan distribution $\pi^{(s)}$.

\subsection{One-unitary marginals of general graphs}
The next class of examples we shall investigate using the techniques developed 
in Section \ref{sec:marginals} is a rather general one. In this section, we consider marginals of graph states with the property that the set $S$ of ``surviving'' subsystems is contained in  one block of the vertex partition $\Pvertex$. 
Note however that we do not impose any conditions on the underlying graph. 

We consider a general graph $\Gamma$ with $k$ nodes and $m$ edges. 
As in Section \ref{sec:graph-states}, to such a graph we associate a random 
pure state $\ket \Psi$ on a $n=2m$-fold tensor product. We are interested in a marginal 
\begin{equation}
\rho_S = \trace_T \ketbra{\Psi}{\Psi} = \trace_{\otimes_{i \in T} \H_i} \ketbra{\Psi}{\Psi},
\label{rhoS}
\end{equation}
with the property that $S = [n]\setminus T \subseteq b$ is contained in a single block $b$
 of the partition $\Pvertex$ defining the vertices of $\Gamma$. Without loss of generality, we can relabel the 
Hilbert spaces $\H_1, \ldots, \H_n$ is such a way that $b = \{1, 2, \ldots, n'\}$, where $n' = |b|$ is the number of subsystem of the ``surviving'' vertex. 

\begin{figure}[htbp]
\centering
\includegraphics{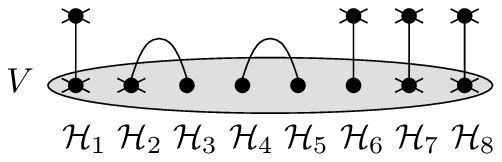}
\caption{Graph illustrating assumptions
  of Theorem \ref{thm:1-unitary-marginal}: all but one vertices are completely traced out.
 The surviving vertex $b$ has  $n'=8$ 
subsystems, out of which $|G|=4$ lead to other vertices,
while the remaining $|F|=n'-|G|$ nodes form closed loops.
Out of these $n'=8$ subsystems, $|T'|=4$ represent subspaces 
which are traced out (crosses),
while $|S|=4$ represent subspaces whose product supports the reduced state $\rho_S$ (full dots). The other vertices of the graph are not represented.}
\label{fig:1-unitary-marginal}
\end{figure}

For a specific example, see Fig. \ref{fig:1-unitary-marginal}, in which $n'=8$. As we shall see, 
the other nodes of the graph $\Gamma$ will not play any role in the statistical properties
 of $\rho_S$, so they are not represented in Figure \ref{fig:1-unitary-marginal}.

The single $n'$-vertex represents a random unitary matrix $U$, which acts on $n'$
 subspaces labeled by the set of indices in the block $b$. As before, $T$ denotes the set of 
subspaces traced out, so the reduced matrix $\rho_S$
lives in the complementary subspace ${\cal H}_S$. Let $T' = [n'] \setminus S = T \cap [n']$ the set of vertices in
 $b$ that are being traced out ($T' \subset T$). We are interested in computing the distribution of the 
random matrix $\rho_S$. Since $\rho_S$ lives on $\H_S \subset \H_b$, the probability distribution of 
the random density matrix $\rho_S$ is invariant with respect to any unitary conjugation on $\H_S$. 

\begin{lemma}\label{lem:unitary-invariance}
The distribution of $\rho_S$ is $\U({\cal H}_S)$-invariant.
\end{lemma}

\begin{figure}[htbp]
\centering
\includegraphics{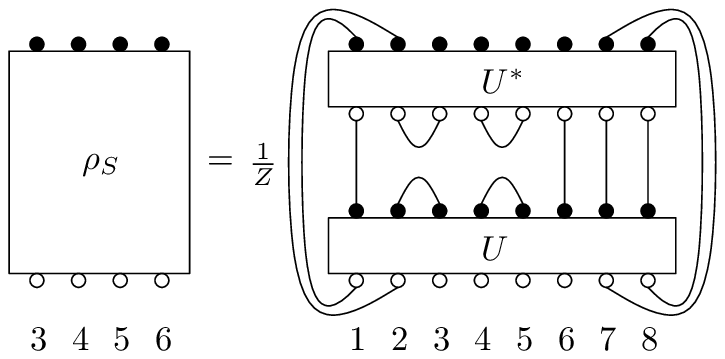}
\caption{The diagram for the random state 
 $\rho_S={\rm Tr}_T|\Psi\rangle \langle \Psi|$
obtained by reduction of the pure state
shown in Fig.\ref{fig:1-unitary-marginal}
}
\label{fig:1-unitary-marginal-gc}
\end{figure}

Two subsets of $[n'] = \{1, \ldots, n'\}$ play an important role in what follows. The set $S$ supporting the reduced density matrix and the set $F$ of bonds which are contained in  $b$: 
\begin{equation}
	F = \bigcup_{ (i,j) \in E \cap [n']} \{i,j\}. 
\end{equation}
In Fig. \ref{fig:1-unitary-marginal}, $S=\{3, 4, 5, 6\}$ and $F=\{2, 3, 4, 5\}$. 
In the graphical notation of \cite{collins-nechita-1}, 
the state $\rho_S$ corresponding to Fig. \ref{fig:1-unitary-marginal} is represented in Figure \ref{fig:1-unitary-marginal-gc}. The unitary blocks corresponding to vertices which were traced out were removed from the diagram, using the \emph{unitary axiom} $UU^* = \I$. Using the box-manipulation rules of \cite{collins-nechita-1}, subsystems in $b$ connected to other vertices in the initial graph are linked now by ``identity wires'' (subsystems $1,6,7,8$ in Figure \ref{fig:1-unitary-marginal-gc}). Notice that although the Hilbert spaces ${\cal H}_i$ with $i\in [n']$ may have different dimensions, we used round-shaped labels for all of them, for obvious practical reasons.

Let the set of bonds complementary to set $F$ be denoted by $G$.
 It 
 consists of $|G|$ nodes which represent bonds connecting nodes of $c$ with nodes belonging to other vertices of $\Gamma$.
In the case shown in Fig. \ref{fig:1-unitary-marginal}
we have  $G=\{1,6,7,8 \}$.
Since the sets $T'$ and $S$, and respectively $F$ and $G$ are complementary with respect to $[n']$, we have $|T'|+|S|=|F|+|G|=n'$

The normalizing constant $1/Z$ appearing in the diagram comes from the different normalizations of 
the input states: $1/(d_i N)$ for every Bell state $\ket{\Phi^+_{ij}}$ associated to an ``internal''
 bond $(i,j) \in F$ and $1/(d_k N)$ for every identity matrix corresponding to an index $k \in G$.
Putting all these contributions together, we have
\begin{equation}
Z = \prod_{i \in F}(d_iN)^{1/2} \times \prod_{j \in G} (d_jN).
\end{equation}

The main result of this section is the following theorem, describing the asymptotic behavior of the 
marginal $\rho_S$. We shall provide two proofs of this result: the first proof is direct, and it uses the graphical calculus in \cite{collins-nechita-1}. 
The second one makes use of Theorem \ref{thm:moments-network} in Section \ref{sec:recipe}.

\begin{theorem}\label{thm:1-unitary-marginal}
For a graph $\Gamma$, let $\ket \Psi$ be the associated random pure state and consider 
\begin{equation}\rho_S = \trace_T \ketbra{\Psi}{\Psi}\end{equation}
a marginal of this random state with the property that $S$ is contained in a single block $b$ of the partition $\Pvertex$ defining the vertices of $\Gamma$. The vertex $b$ contains $n'$ nodes, out of which $|G|$ are connected to other vertices. The partial trace above is taken over $|T|$ subspaces of joint  dimension
$N^{|T|}d_T$. The reduced density matrix $\rho_S$,
supported by the complementary $|S|$ subspaces of dimension $d_S N^{|S|}$,
is asymptotically ($N \to \infty$) characterized by the following behavior
\begin{enumerate}
\item[(I)] If $|S| < |T'|+|G|$,
 the state $\rho_S$ converges in moments to the maximally mixed state in ${\cal H}_S$.
\item[(II)] If $|S| = |T'|+|G|$, then the rescaled operator
    $d_{T'} d_{G} N^{|S|} \rho_S$ converges in moments to a 
  free Poisson distribution $\pi^{(1)}_c$ (see equation \eqref{eq:freepoiss}) of parameter $c={d_{T'} d_{G}} / {d_S}$.
\item[(III)] If $|S| > |T'|+|G|$, the reduced state $\rho_S$ has rank $d_{T'}d_{G} N^{|T'| + |G|}$. If $\tilde \rho_S$ is the restriction of $\rho_S$ to its support, then $\tilde \rho_S$
 converges in moments to the maximally mixed state in  $\H_{T'} \otimes \H_G$. The support of $\rho_S$ is a $d_{T'}d_{G} N^{|T'| + |G|}$ dimensional Haar random subspace of $\H_S$.
\label{th44}
\end{enumerate}
\end{theorem}

\begin{proof}
Using the unitary invariance, it is easy to see that one can replace the tensor products of maximally 
entangled spaces by any rank-one projector. Using this trick and collecting all the vector spaces corresponding to the
sets $S$, $T'$, $F$, $G$, one obtains the simplified diagram for $\rho_S$ 
depicted in Figure \ref{fig:1-unitary-marginal-gc-simple}.

\begin{figure}[htbp]
\centering
\includegraphics{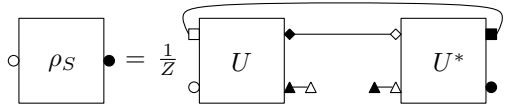}
\caption{Simplified diagram for reduced state $\rho_S$ corresponding to Fig.
\ref{fig:1-unitary-marginal-gc}.
Round-shaped labels correspond to the subspace 
${\cal H}_S$ of dimension $d_S N^{|S|}$, square-shaped labels to 
the traced out subspace ${\cal H}_T$ of dimension 
$d_{T} N^{|T|}$, triangles to $d_F N^{|F|}$ and 
diamonds to $d_{G} N^{|G|}$. The normalization constant
reads $Z = d_{G} N^{|G|}$.
}
\label{fig:1-unitary-marginal-gc-simple}
\end{figure}

Following Theorem \ref{thm:moments},
one can compute the moments of the random matrix $\rho_S$ 
using the Weingarten function $\Wg(k, \alpha)$ and the
graphical calculus:

\begin{equation}
\E \trace(\rho_S^p) = Z^{-p} \sum_{\alpha, \beta \in \S_p}
 \left[ d_S N^{|S|}\right]^{\#(\gamma^{-1} \alpha)} \left[ d_{T} N^{|T'|} \right]^{\#\alpha}
 \left[ d_{G} N^{|G|} \right]^{\#\beta} \Wg(d_{[n']}N^{n'},\alpha^{-1} \beta).
\label{etrace}
\end{equation}

Since we are interested in the asymptotical regime $N \to \iy$, we investigate the power of $N$ in the sum above:
\begin{equation}N^{-p|G|} N^{|S|(p-|\gamma^{-1}\alpha|)} N^{|T'|(p-|\alpha|)} N^{|G|(p-|\beta|)} N^{n'(-p-|\alpha^{-1}\beta|)}.\end{equation}

Using the equality $n' =|S| + |T'|$, we find that the exponent of $N$ that one wants to maximize in order to obtain the dominating terms is 
\begin{equation}\label{eq:maxim-problem}
-(|S||\gamma^{-1}\alpha| + |T'||\alpha| + |G||\beta|+n'|\alpha^{-1}\beta|).
\end{equation}
Using the triangular identities 
\begin{align}
\label{eq:triangular-1}|\gamma^{-1}\alpha| +
 |\alpha^{-1}\beta| &\geq |\gamma^{-1}\beta|\\
\label{eq:triangular-2}|\alpha| + |\alpha^{-1}\beta| &\geq |\beta|
\end{align}
$|S|$ times and $|T'|$ times respectively, we look at the (a priori) weaker minimization problem over $\beta \in \S_p$
\begin{equation}\label{eq:minimization-problem-beta}
\text{minimize } \quad f(\beta) = |S||\gamma^{-1}\beta| + (|T'|+|G|) |\beta|.
\end{equation}

Before solving this minimization problem, let us consider the extremal cases where $S = \emptyset$ or $T' = \emptyset$. If $S = \emptyset$, then every subsystem of $\ket \Psi$ is traced out, and the reduced state is degenerate: $\rho_S = 1 \in \M_1(\C)$. If $T' = \emptyset$, then, for all $N$, $\rho_S$ is distributed as a Haar random projector of rank $d_G N^{|G|}$ in $\M_{d_S N^{|S|}}(\C)$. From now on, we assume that $|S|, |T'| \geq 1$. We go back to the minimization problem \eqref{eq:minimization-problem-beta}, which is easily solved:
\begin{equation}\min_{\beta \in \S_p} f(\beta) = (p-1) \cdot \min\{|S|, |T'|+|G|\}.\end{equation}
The set of permutations $\beta$ which reach the above minimum is as follows:
\begin{equation}\label{eq:argmin-beta}
\argmin f(\beta) = \begin{cases}
\{\id\} \quad &\text{ if } |S| < |T'|+|G|;\\
\{\beta \; |\; \id \to \beta \to \gamma \text{ geodesic}\} \quad &\text{ if } |S| = |T'|+|G|;\\
\{\gamma\} \quad &\text{ if } |S| > |T'|+|G|.
\end{cases}
\end{equation}

\bigskip

Having solved the minimization problem \eqref{eq:minimization-problem-beta}, let us now return to the original maximization problem in $\alpha$ and $\beta$, equation \eqref{eq:maxim-problem}. Since (except in the trivial case where $S$ or $T'$ is the empty set) both triangular inequalities \eqref{eq:triangular-1} and \eqref{eq:triangular-2} have been used at least once, and since the only permutation $\alpha$ which saturates both inequalities at fixed $\beta$ is $\alpha=\beta$ (see equations \eqref{eq:alpha_i-beta_i} - \eqref{eq:alpha_i-beta_i-chain}), the dominating terms in the Weingarten sum are those for which $\alpha=\beta$ and $\beta$ is optimal as in equation \eqref{eq:argmin-beta}. We discuss now each case in equation \eqref{eq:argmin-beta} separately:

\begin{enumerate}
\item Case (I): $|S| < |T'|+|G|$. The dominating term is given by $\alpha = \beta = \id$. We get
an expression for the expectation value of the power of the traces of the mixed state under consideration.

\begin{equation}\label{eq:1-unitary-case-I}
\E \trace(\rho_S^p) \sim N^{|S|(1-p)} d_{G}^{-p} d_S d_{T'}^p d_{G}^p d_{[n']}^{-p} = \left( d_S N^{|S|} \right) ^{1-p}.
\end{equation}
Thus, $\rho_S$ behaves as a maximally mixed state on the $d_S N^{|S|}$-dimensional subspace ${\cal H}_S$.

\item Case (II): $|S| = |T'|+|G|$. Here, dominating terms are indexed by geodesic permutations $\id \to \alpha = \beta \to \gamma$. We get
\begin{equation}\label{eq:1-unitary-case-II}
\begin{split}
\E \trace(\rho_S^p) &\sim N^{|S|(1-p)} d_{G}^{-p} \sum_{\id \to \alpha \to \gamma} d_S^{\#(\gamma^{-1}\alpha)} d_{T'}^{\#\alpha} d_{G}^{\#\alpha} d_{[n']}^{-p} \\
&= \left( d_S N^{|S|} \right) ^{1-p} \left( \frac{d_{T'} d_{G}}{d_S} \right)^{-p} \sum_{\id \to \alpha \to \gamma}\left( \frac{d_{T'} d_{G}}{d_S} \right)^{\#\alpha}.
\end{split}
\end{equation}
We conclude that the (renormalized) matrix $ d_{T'} d_{G}N^{|S|} \rho_S$ converges in moments to a free Poisson distribution $\pi^{(1)}_c$ of parameter $c={d_{T'} d_{G}} / {d_S}$ (see equation \eqref{eq:freepoiss}).

\item Case (III): $|S| > |T'|+|G|$. Here, the dominating term is the one with $\alpha = \beta = \gamma$. The asymptotic moments of $\rho_S$ are given by
\begin{equation}
\E \trace(\rho_S^p) \sim  \left( d_{T'}d_{G} N^{|T'| + |G|} \right) ^{1-p}.
\end{equation}
We conclude that $\rho_S$ behaves like a maximally mixed state of dimension $d_{T'}d_{G} N^{|T'| + |G|}$. 
Notice that this is the maximum rank of $\rho_S$ in this case, since we are partial tracing over a space  ${\cal H}_{T'}$ of dimension $d_{T'}N^{|T'|}$ a state of rank $d_{G}N^{|G|}$. Hence, $\rho_S$, restricted to its support, converges in moments to a maximally mixed state. The distribution of the support of $\rho_S$ is Haar, by the unitary invariance Lemma \ref{lem:unitary-invariance}.
\end{enumerate}
\end{proof}

We shall now present a second proof of this result, using the flow network method of
 Theorems \ref{thm:moments} and \ref{thm:moments-network-N}.
 We invite the reader to judge the efficiency of the network method in finding the
 set of optimal permutations.

\begin{proof}
The network associated to a one-unitary marginal is presented in Figure\ref{fig:1-unitary-marginal-network}. Since vertices $\beta_2, \ldots, \beta_k$ are connected only to the source, one should have $\beta_2 = \cdots = \beta_k = \id$ in the end. Hence, we can just ignore these vertices: they can only be used to send a maximum flow of $|G|$ from $\id$ to $\beta_1$. The updated, simplified reduced network is represented in Figure \ref{fig:1-unitary-marginal-network-equiv}.

\begin{figure}[htbp]
\centering
\subfigure[]{\label{fig:1-unitary-marginal-network}\includegraphics{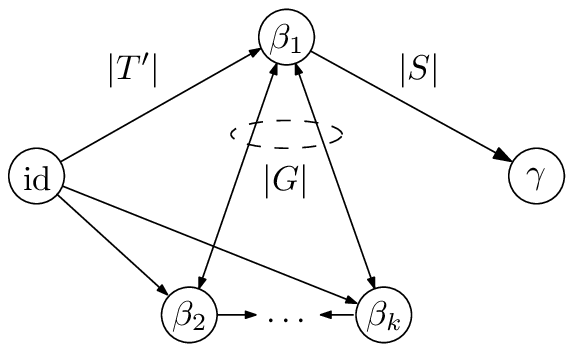}} \quad\quad
\subfigure[]{\label{fig:1-unitary-marginal-network-equiv}\includegraphics{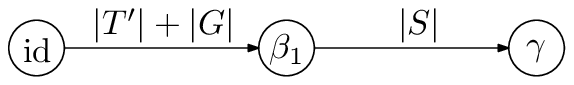}}
\caption{Network associated to a one-unitary marginal (a) and its equivalent form (b) obtained by ignoring the vertices $\beta_2, \ldots, \beta_k$, which are traced out entirely.}
\end{figure}

The maximum flow in the above network is $X = \min\{|S|, |T'|+|G|\}$ and the residual network depends on which of $|S|$ and $|T'|+|G|$ is greater, see Figure \ref{fig:1-unitary-marginal-network-resid}. 

\begin{figure}[htbp]
\centering
\includegraphics{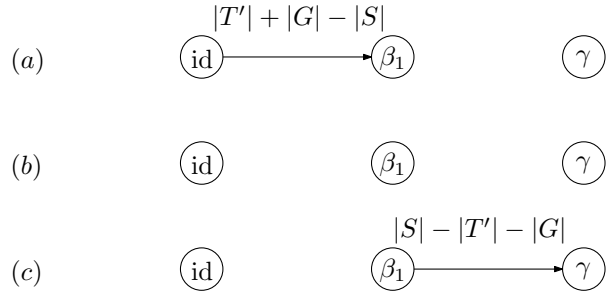}
\caption{Residual networks for one-unitary marginals, in the following cases: (a) $|S|<|T'|+|G|$ ; (b)  $|S|=|T'|+|G|$ ; (c)  $|S|>|T'|+|G|$}
\label{fig:1-unitary-marginal-network-resid}
\end{figure}

In the first case, where $|S|<|T'|+|G|$, the permutation $\beta_1$ is connected to the source by an edge of positive capacity $|T'|+|G|-|S| >0$ and thus $\beta_1 = \id$. In this case, we conclude that all permutations $\alpha_i, \beta_i$ are equal to the identity $\id$. Hence, the asymptotic moments of the one-unitary marginal $\rho_S$ are given by plugging into equation \eqref{eq:moment_general_Wg} the values of the optimal permutations:
\begin{equation}
\begin{split}
\E \trace(\rho_S^p) &\sim Z^{-p}  \left( d_{S}N^{|S|} \right)\prod_{i=1}^k \left( d_{T_i}N^{|T_i|} \right)^p \\
&\cdot \prod_{1 \leq i < j \leq k} \left( d_{E_{ij}}N^{|E_{ij}|} \right)^p \prod_{i=1}^k (d_{C_i}N^{|C_i|})^{-p}.
\end{split}
\end{equation}

In the above formula, we have $S=S_1$, $T' = T_1$ and $G = \sqcup_{j=2}^k E_{1j}$. After simplifying the factors in the products with the factors in $Z$, we are left with 
\begin{equation}\E \trace(\rho_S^p) \sim \left( d_{S}N^{|S|} \right)^{1-p},\end{equation}
which allows to conclude and is consistent with equation \eqref{eq:1-unitary-case-I}. The third case, where $|S|>|T'|+|G|$ can be proved in an analogous way.

Let us consider now the ``critical'' case, $|S|=|T'|+|G|$. Since the residual network is empty (it has no edges), there is no constraint on the permutation $\beta_1$, beside that it is required to be geodesic. The permutations $\alpha_i, \beta_i$ which contribute asymptotically are those such that
\begin{equation}
\begin{cases}
&\alpha_1 = \beta_1 \\
&\hat 0_p \leq [\beta_1] \leq \hat 1_p \\
&\alpha_i = \beta_i = \id \quad \forall i=2,3,\ldots, k
\end{cases}
\end{equation}

Plugging these values into equation \eqref{eq:moment_general_Wg}, we recover the final result in equation \eqref{eq:1-unitary-case-II}, and the proof is complete.
\end{proof}

\subsection{Star graphs}

\begin{figure}[htbp]
\centering
\subfigure[]{\label{fig:star-graph-simplified}\includegraphics{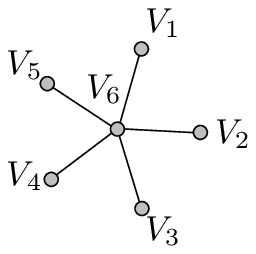}} \quad\quad
\subfigure[]{\label{fig:star-graph}\includegraphics{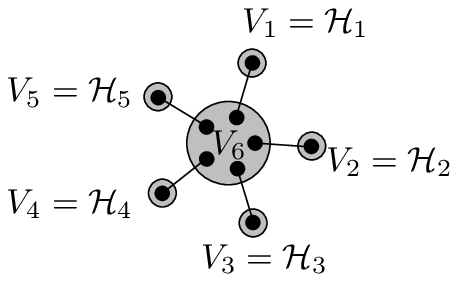}}
\caption{A $5$-star graph state in the simplified and usual graphical notation}
\label{fig:5-star-graph}
\end{figure}

In this section we investigate random quantum states associated to \emph{star graphs}. A $m$-star graph is a graph with $k=m+1$ vertices $V_1, V_2, \ldots, V_m, V_{m+1}$ and edges $\{(j, m+1)\}_{j=1}^m$ (see Figure \ref{fig:star-graph-simplified}). For obvious graphical reasons, the vertices $V_1, V_2, \ldots, V_m$ shall be called ``satellites'' and the distinguished vertex $V_{m+1}$ shall be called the ``center'' of the graph. The central vertex is a tensor product of $m$ subspaces
\begin{equation}
W_{m+1} = \otimes_{i=m+1}^{2m} \H_i,
\end{equation}
while the satellites have only one subsystem: $W_i = \H_i$, for $i=1, \ldots, m$.
Edges correspond to entangled states between $\H_i$ and $\H_{m+i}$ for $i=1, \ldots, m$. We shall consider the simplified situation where $d_i = 1$, and thus $\H_i \isom \C^N$ for all $i=1, \ldots, 2m$. The graph state $\ket{\Psi_\text{star}} \in (\C^N)^{\otimes 10}$ associated to an $m=5$-star is depicted in figure \ref{fig:star-graph}. Notice that the Hilbert spaces corresponding to the vertices are given by $W_{m+1} = (\C^N)^{\otimes m}$ and $W_i = \C^N$ for $i=1, 2 \ldots, m$. We have
\begin{equation}
\ket{\Psi_\text{star}} = \left[ \left (\bigotimes_{i=1}^m U_i\right) \otimes U_{m+1}\right]\left(\bigotimes_{j=1}^m \ket{\Phi^+_{j, m+j}}\right).
\end{equation}
The unitary matrices $U_1, \ldots, U_{m+1}$ are independent Haar-distributed random unitary matrices $U_i \in \U(N)$, $i=1, \ldots, m$ and $U_{m+1} \in \U(N^m)$.

Before looking at general marginals of the star graph state, some simplifications can be made at this point. First, notice that one can assume $U_1 = \cdots = U_m = \I$. This follows from the fact that the unitary transformation $U_1 \otimes \cdots \otimes U_m$ can be ``absorbed'' into $U_{m+1}$; for a Bell state $\ket{\Phi^+}$ and an unitary transformation $U$, 
\begin{equation}(U \otimes \I) \ketbra{\Phi^+}{\Phi^+}(U \otimes \I)^* = (\I \otimes \ol U) \ketbra{\Phi^+}{\Phi^+}(\I \otimes \ol U)^*,\end{equation}
which is a consequence of the well-known fact that the commutant of the group $G$ of local unitaries $G=\{U \otimes \ol U \; |\; U \in \U(N)\}$ is spanned by the identity and the maximally entangled state: $G' = \Span\{\I, \ketbra{\Phi^+}{\Phi^+}\}$. 

Second, since the system is invariant with respect to permutations of the satellites $1, \ldots, m$, a general marginal is specified by a coupe $(s,t)$ of integers from $[m]$, where $m-s$ is the number of satellites that have been traced out and $m-t$ is the number of traced particles in the central block $0$. The resulting marginal is $\rho^{(s,t)} \in (\C^N)^{\otimes(s+t)}$: $s$ copies of $\C^N$ have ``survived'' in the satellites and $t$ copies in the central vertex. Let us make a remark about the rank of the matrix $\rho^{(s,t)}$. Obviously, we have $\rk \rho^{(s,t)} \leq s+t$. Moreover, since we are partial tracing a pure state over $(m-s) + (m-t) = 2m-(s+t)$ copies of $\C^N$, we also have $\rk \rho^{(s,t)} \leq 2m-(s+t)$. In conclusion, we obtain the bound $\rk \rho^{(s,t)} \leq \min(s+t, 2m-(s+t))$.

\begin{figure}[htbp]
\centering
\subfigure[]{\includegraphics{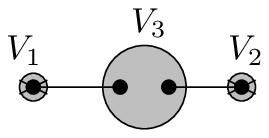}} \quad\quad
\subfigure[]{\includegraphics{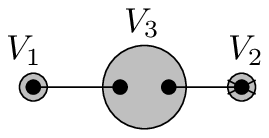}} \quad\quad
\subfigure[]{\includegraphics{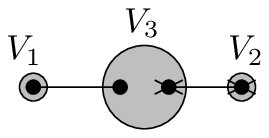}}
\caption{Graphs corresponding to the marginals $\rho^{(0,2)}$, $\rho^{(1,2)}$ and $\rho^{(1,1)}$ respectively.}
\label{fig:2-star-examples}
\end{figure}

\begin{figure}[htbp]
\centering
\subfigure[]{\label{fig:2-star-gc-a}\includegraphics{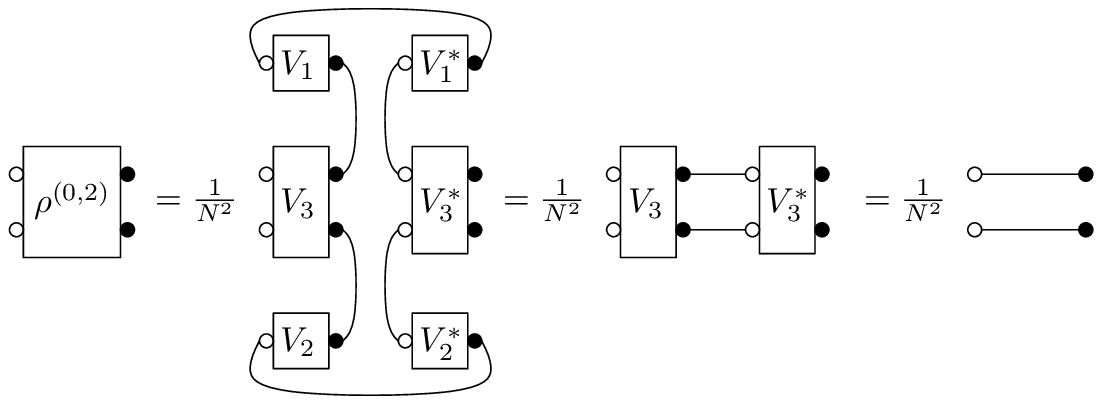}} \\
\subfigure[]{\includegraphics{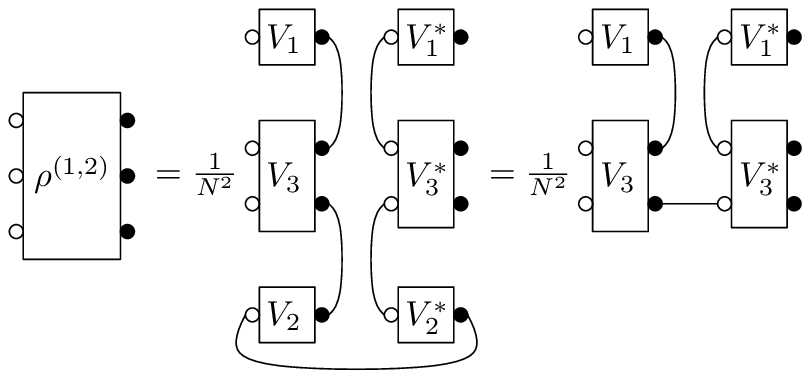}} \\
\subfigure[]{\includegraphics{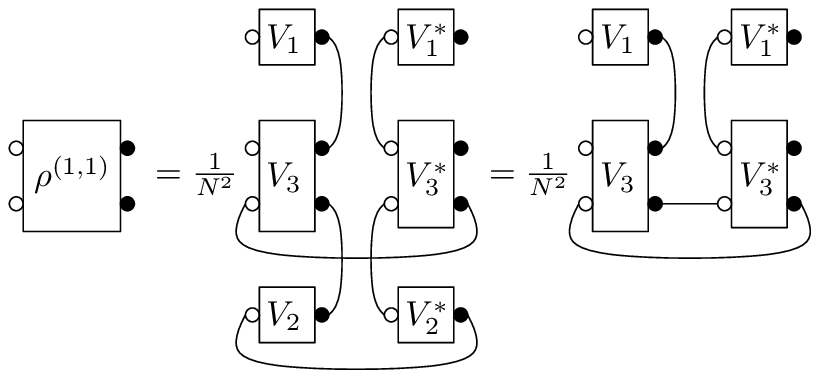}}
\caption{Graphical notation for $\rho^{(0,2)}$, $\rho^{(1,2)}$ and $\rho^{(1,1)}$.}
\label{fig:2-star-gc}
\end{figure}

 We shall look in detail at three examples, for $m=2$-star graphs: $(s=0, t=2)$,  $(s=1, t=0)$ and $(s=t=1)$. The graphs corresponding to these marginals are represented in Figure \ref{fig:2-star-examples}. 
We use the graphical formalism of \cite{collins-nechita-1} to represent the (random) density matrices $\rho^{(0,2)}$, $\rho^{(1,2)}$ and $\rho^{(1,1)}$ in Figure \ref{fig:2-star-gc}.

In Figure \ref{fig:2-star-gc-a}, we use the rules for manipulating boxes in the graphical formalism to show that $\rho^{(0,2)} = \frac{1}{N^2} \I_{N^2}$. Actually, we start by collapsing the $V_1$ and the $V_2$ boxes (and their adjoints) and then we use the same rule to collapse $V_3$ and $V_3^*$. These considerations can be easily generalized to obtain the following lemma. 

\begin{lemma}
For all $s,t \in [m]$, the marginals $\rho^{(0,t)}$ and $\rho^{(s,0)}$ are given by the maximally mixed states in $\frac{1}{N^t} \I_{N^t} \in \M_{N^t}(\C)$ and $\frac{1}{N^s} \I_{N^s} \in \M_{N^s}(\C)$ respectively. 
\end{lemma}

\begin{figure}[htbp]
\centering
\subfigure[]{\label{fig:2-star-1-2-network}\includegraphics{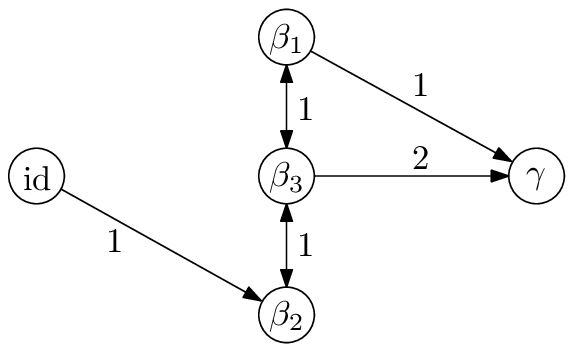}} \quad\quad
\subfigure[]{\label{fig:2-star-1-1-network}\includegraphics{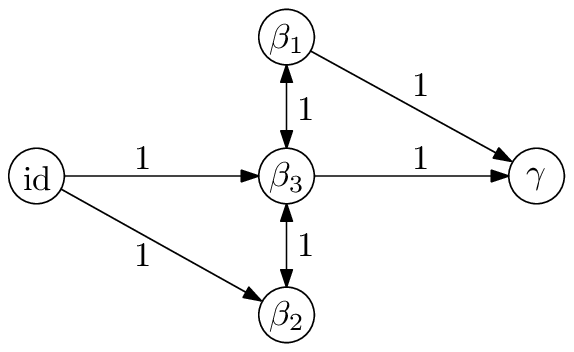}}
\caption{Networks associated to the minimization problems for $\rho^{(1,2)}$-(a) and $\rho^{(1,1)}$-(b).}
\label{fig:2-star-network}
\end{figure}

Let us analyze the marginal $\rho^{(1,2)}$. The network corresponding to the minimization problem associated to this marginal is represented in Figure \ref{fig:2-star-1-2-network}.

Since the vertex $\beta_2$ is connected only to the source, and the other two vertices $\beta_3$ and $\beta_1$ are only connected to the sink, the solution for the minimization problem is given by 
\begin{equation}\begin{cases}
\alpha_3 = \beta_3 = \alpha_1 = \beta_1 = \gamma, \\
\alpha_2 = \beta_2 = \id.
\end{cases}\end{equation}
The maximum flow in the network is 1, and thus the asymptotic moments of the marginal are given by Theorem \ref{thm:moments-network-N}: 
\begin{equation}\frac{1}{N}\E \trace(\rho^{(1,2)})^p = \frac{1}{N^p} + o(N^{-p}).\end{equation}
In other words, the rescaled random density matrix $N\rho^{(1,2)}$ converges in distribution to the Dirac mass $\delta_1$. In physical terms, the random matrix $\rho^{(1,2)}$ (which has rank at most $N$) behaves asymptotically as a (normalized to unit trace) projector on a subspace of dimension $N$ of $\C^{N^3}$.

We finish our case studies by the marginal $\rho^{(1,1)}$ of a $2$-star graph. The flow network associated to $\rho^{(1,1)}$ is depicted in Figure \ref{fig:2-star-1-1-network}. The maximum flow in this network is 2: a unit of flow is sent through the augmenting paths $\id \to \beta_3 \to \gamma$ and $\id \to \beta_2 \to \beta_3 \to \beta_1 \to \gamma$. The residual network is empty and thus, the geodesic permutations $\beta_{1,2,3}$ which achieve the minimum should satisfy $\beta_2 \leq \beta_3 \leq \beta_1$. Since $\beta_1$ is connected only to $\gamma$ and $\beta_2$ is connected only to the source $\id$, the sum over the M\"{o}bius functions constrain $\beta_1$ and $\beta_2$:
\begin{equation}\begin{cases}
\alpha_1 = \beta_1 = \gamma, \\
\alpha_2 = \beta_2 = \id, \\
\hat 0_p \leq [\alpha_3 = \beta_3] \leq \hat 1_p.
\end{cases}\end{equation}

It follows that the asymptotic moments of the marginal $\rho^{(1,1)}$ are given by
\begin{equation}\frac{1}{N}\E \trace(\rho^{(1,1)})^p = \frac{1}{N^{2p}}\sum_{\pi \in NC(p)}1 + o(N^{-2p}) = N^{-2p} \frac{1}{p+1} \binom{2p}{p} + o(N^{-2p}).\end{equation}
We conclude that the renormalized density matrix $N^2 \rho^{(1,1)}$ converges in distribution to a free Poisson distribution (of parameter $c=1$) $\pi^{(1)}$.

The above discussion can be easily generalized to obtain the following general theorem. 

\begin{theorem}
The marginal $\rho^{(s,t)} \in \M_{N^{s+t}}(\C)$ obtained by partial tracing $m-s$ satellites and $m-t$ central systems of a random star graph state $\ket{\Psi_\text{star}}$ has the following asymptotic behavior:
\begin{enumerate}
\item If $s=0$ (resp. $t=0$), then, for all $N$,
\begin{equation} \rho^{(0,t)} = \frac{1}{N^t} \I_{N^t} \quad (\text{resp. }  \rho^{(s,0)} = \frac{1}{N^s} \I_{N^s}).\end{equation}
\item If $s+t < m$, then $N^{s+t} \rho^{(s,t)}$ converges in distribution to $\delta_1$.
\item If $s+t > m$, then $\rho^{(s,t)}$ has rank $2m - (s+t)$. If $\tilde\rho^{(s,t)}$ is the restriction of $\rho^{(s,t)}$ to its support, then $N^{2m-(s+t)} \tilde \rho^{(s,t)}$ converges in distribution to $\delta_1$.
\item If $s+t=m$ and $s,t \neq 0$ then $N^m \rho^{(s,t)}$ converges in distribution to a Free Poisson distribution $\pi^{(1)}$.
\end{enumerate}
Moreover, the average von Neumann entropy and the average purity of the marginals $\rho^{(s,t)}$ are given in the following table:
\begin{table}[h]
	\centering
		\begin{tabular}{|r|c|c|}
			\hline
			Parameters & von Neumann entropy $\E H(\rho^{(s,t)})$ & Purity $\E \trace\left((\rho^{(s,t)})^2\right)$\\
			\hline\hline
			$s=0$ (resp. $t=0$) & $t \log N$ (resp. $s \log N$)& $N^{-t}$ (resp. $N^{-s}$) \\ \hline
			$s+t < m$ & $(s+t) \log N + o(N^{-(s+t)})$ & $N^{-(s+t)} + o(N^{-(s+t)})$ \\ \hline
			$s+t > m$ & $(2m - s - t) \log N + o(N^{-(2m - s-t)})$ & $N^{-(2m - s - t)} + o(N^{-(2m - s-t)})$ \\ \hline
			$s+t = m$; $s,t \neq 0$ & $m \log N - 1/2 + o(1)$ & $2N^{-m} + o(N^{-m})$ \\ \hline
		\end{tabular}
	\caption{von Neumann entropy and purity for marginals of star graph states.}
	\label{tab:entropy-purity-star}
\end{table}
\end{theorem}

\subsection{Cycle graphs}
In this section we look at random quantum pure states associated to cycle graphs. The asymptotic eigenvalue distributions of marginals of such graphs will be characterized in terms of the subset of traced systems. The set of possible limit measures one can obtain in this situation is fairly rich: for any classical multiplicative convolution of measure from the Fuss-Catalan family, one can construct a cycle graph having this measure as the limit eigenvalue distribution.

\begin{figure}[htbp]
\centering
\subfigure[]{\label{fig:4-cycle-graph-simplified}\includegraphics{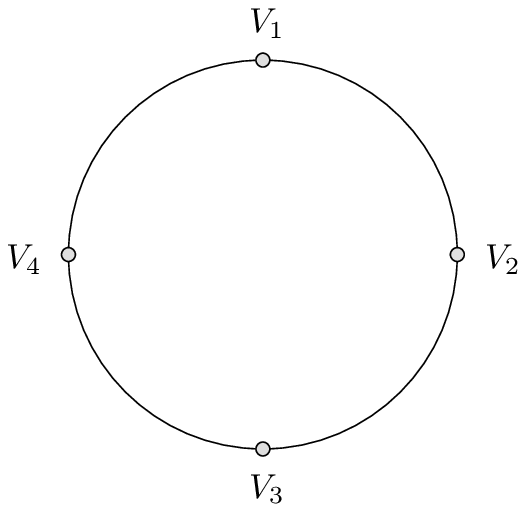}} \quad\quad
\subfigure[]{\label{fig:4-cycle-graph}\includegraphics{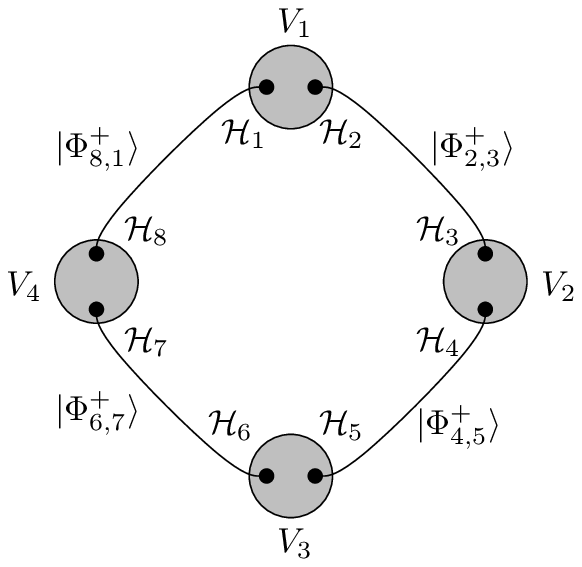}}
\caption{A $4$-cycle graph state in the simplified and usual graphical notation}
\label{fig:4-cycle-graph-tot}
\end{figure}

An $m$-cycle is a graph having $k=m$ vertices and $m$ edges connecting the vertices in a cyclic way. The simplified and detailed representations of a $m=4$-cycle are given in Figures \ref{fig:4-cycle-graph-simplified} and \ref{fig:4-cycle-graph}. Each vertex of the graph contains two subsystems, thus the whole graph has $n=2m$ subsystems. To keep notation simple, we shall assume that $d_i = 1$ for all $i=1, \ldots, 2m$. Hence, the vector spaces $\H_1, \ldots, \H_{2m}$ are isomorphic to $\C^N$. Vertices are vector spaces
\begin{equation}W_i  = \H_{2i-1} \otimes \H_{2i} \isom \C^{N^2}.\end{equation}
The $m$ edges of the graph correspond to maximally entangled states $\ket{\Phi^+_{2j, 2j+1}} \in \H_{2i} \otimes \H_{2i+1} \isom \C^N \otimes \C^N$ (with the obvious convention $2m+1 = 1$):
\begin{equation}
\ket{\Psi_\text{cycle}} = \left[ \bigotimes_{i=1}^m U_i \right]\left(\bigotimes_{j=1}^m \ket{\Phi^+_{2j, 2j + 1}}\right),
\end{equation}
where the random unitary operator $U_i$ acts on $W_i = \H_{2i-1} \otimes \H_{2i}$.

We are interested in the eigenvalue statistics of the marginals
\begin{equation}\rho_S = \trace_T \ketbra{\Psi_\text{cycle}}{\Psi_\text{cycle}},\end{equation}
parameterized by partitions $\Ptrace = \{S, T\}$ of $[n]$. The number of systems being traced out in each vertex can be 0, 1 or 2, so we can classify the $m$ vertices of a cycle in three classes:
\begin{enumerate}
\item Vertices of type ``S'', where nothing is traced out;
\item Vertices of type ``R'', where exactly one system is traced out;
\item Vertices of type ``T'', where both systems are traced out.
\end{enumerate}
Notice that we are implicitly using the fact that in a vertex with systems of equal dimension, the exact indices of the systems being traced out are not relevant; in this way, the definition of type ``R'' vertices is not ambiguous.

\begin{figure}[htbp]
\centering
\subfigure[]{\label{fig:cycle-graph-marginal}\includegraphics{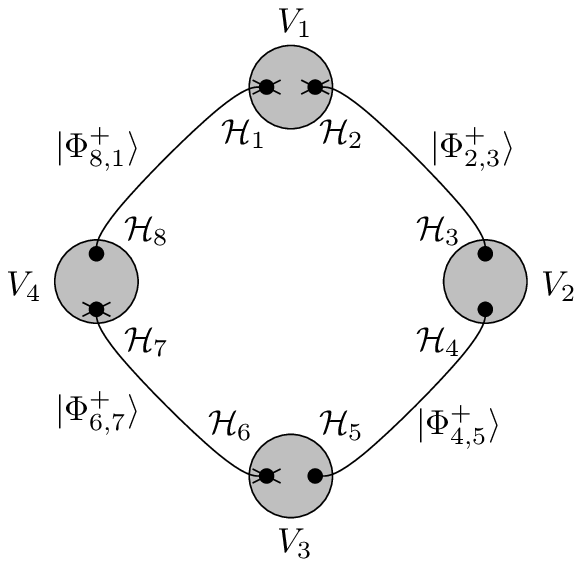}} \quad\quad
\subfigure[]{\label{fig:cycle-graph-marginal-network}\includegraphics{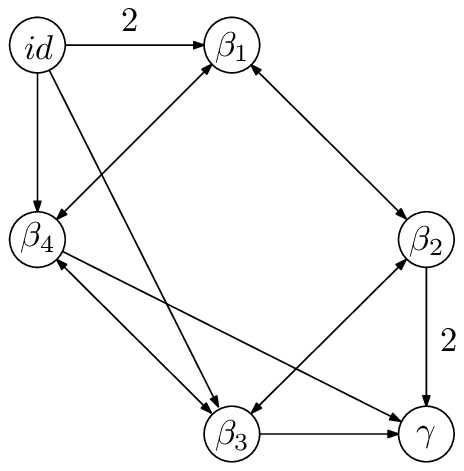}}
\caption{(a) Marginal of a $4$-cycle graph with one ``T'' vertex ($V_1$), one ``S'' vertex ($V_2$) and two ``R'' vertices ($V_3$ and $V_4$). (b) The network associated to the marginal; edges with no labels have unit capacity.}
\end{figure}

In Figure \ref{fig:cycle-graph-marginal} we consider a marginal of the $m=4$-cycle, with $S = \{3,4,5,8\}$ and $T = \{1,2,6,7\}$. Vertex $V_1$ is of type ``T'', $V_2$ is of type ``S'' and vertices $V_3$ and $V_4$ are of type ``R''. We shall work out this example in detail, since it is the simples situation where the Fuss-Catalan distribution of order 2, $\pi^{(2)}$ emerges. 

The network associated to the marginal $\rho_S$ is represented in Figure \ref{fig:cycle-graph-marginal-network}.  The maximum flow problem on the network has an unique solution which can be computed in the following way. First, send one unit of flow through the ``R'' vertices $\id \to \beta_3 \to \gamma$, $\id \to \beta_4 \to \gamma$. The only remaining way of sending flow from the source $\id$ to the sink $\gamma$ is through the arcs $\id \to \beta_1 \to \beta_2 \to \gamma$ and $\id \to \beta_1 \to \beta_4 \to \beta_3 \to \beta_2 \to \gamma$. In this way, two additional units of flow can be sent from $\id$ to $\gamma$. After sending 4 units of flow (2 using ``R'' vertices and 2 using ``TS'' and ``TRRS'' arcs), the residual network is empty and the solution to the maximal flow problem is thus $X=4$. 

The set of geodesic permutations $\beta_1, \ldots, \beta_4$ which achieve the maximum flow is constrained only by the inequality $[\beta_1] \leq [\beta_4] \leq [\beta_3] \leq [\beta_2]$ (recall that for a permutation $\beta$, $[\beta]$ denotes the partition induced by the cycle structure of $\beta$). The fact that $\beta_1$ (or $V_1$) was only connected to the source (i.e. it was completely traced out) imposes the additional constraint $\beta_1 = \id$. Similarly, since $\beta_2$ is only connected to the sink (i.e. none of its subsystems was traced out), it must be that $\beta_2 = \gamma$. It follows that permutations $\alpha_i, \beta_i$ which achieve the minimum in equation \eqref{eq:minimization-problem-beta} are those that verify 
\begin{align*}
\begin{cases}
&\alpha_1 = \beta_1 = \id \\
&\alpha_2 = \beta_2\\
&\alpha_3 = \beta_3\\
&\alpha_4 = \beta_4 = \gamma\\
&[\beta_4] \leq [\beta_3]
\end{cases}
\end{align*}

Using Theorem \ref{thm:moments-network-N}, we conclude that the moments of the above marginals are given by the following expression
\begin{equation}\label{eq:moments-cycle-example}
\lim_{N \to \iy} N^{4(p-1)}\E \trace \rho_S^p = |\{\sigma_1, \sigma_2 \in NC(p) \; | \; \sigma_1 \leq \sigma_2\}|.
\end{equation}
The combinatorial factor is the above equation is the number of chains of length 2 in $NC(p)$, the Fuss-Catalan number $FC^{(2)}_p$, see also equation \eqref{eq:def-comb-FC}. We conclude that the random density matrix $N^4 \rho_S$ converges in distribution to the Fuss-Catalan probability measure of order 2, $\pi^{(2)}$.

One can easily generalize the previous considerations to all $m$ and arbitrary marginals to obtain the following the main result of this section, a complete characterization of the limiting eigenvalue statistics for marginals of random cycle graph states. 

\begin{theorem}\label{thm:cycle-limit}
The asymptotic moments of the rescaled random density marginal $\rho_S$ are products of Fuss-Catalan numbers corresponding to ``TR$\cdots$RS'' arcs:
\begin{equation}\lim_{N \to \iy} \E N^{(k_R + |\mathcal A|)(p-1)} \trace \rho_S^p = \prod_{a \in \mathcal A} FC^{(|a|)}_p= \prod_{a \in \mathcal A} \frac{1}{|a|p +1} \binom{(|a|+1)p}{p},\end{equation}
where $\mathcal A$ is the set of all ``TR$\cdots$RS'' arcs in the cycle graph, $|a|$ is the length of an arc $a \in \mathcal A$ and $k_R$ is the number of type ``R'' vertices.
The empirical eigenvalue distribution of a rescaled version of $\rho_S$ converges to a classical multiplicative convolution product of Fuss-Catalan probability distributions $\times_{a \in \mathcal A} \pi^{(|a|)}$. Moreover, the average von Neumann entropy of the random density matrix $\rho_S$ is given by
\begin{equation}
\E H(\rho_S) = (k_R + |\mathcal A|) \log N - \sum_{a \in \mathcal A} \sum_{j=2}^{|a|+1} \frac{1}{j} + o(1).
\end{equation}
\end{theorem}
\begin{proof}
For a general marginal $\rho_S$ of a random cycle graph state $\ket{\Psi_\text{cycle}}$, let $k_{R,S,T}$ be the respective numbers of ``R'',''S'' and ``T'' vertices in the marginal; one has $k_R + k_S + k_T = k = m$. Let also $\mathcal A$ be the set of all ``TR$\cdots$RS'' arcs. An element $a \in \mathcal A$ is a sequence of $|a|$ consecutive vertices $V_{a_0}, V_{a_1}, \ldots, V_{a_{|a|+1}}$ such that $V_{a_0}$ is an ``T'' type vertex, $V_{a_{|a|+1}}$ is an ``S'' type vertex and all intermediate elements $V_{a_1}, \ldots, V_{a_{|A|}}$ are of type ``R'' (we consider only non-empty arcs, hence $|a|>0$). Arguing as we did in the previous example, the maximum flow in the general case is equal to $k_R + |\mathcal A|$: one sends 1 unit of flow through each ``R'' vertex and 1 unit through each ``TR$\cdots$RS'' arc $a \in \mathcal A$. The set of (geodesic) permutations $\{\alpha_i, \beta_i\}_{i=1}^m$ which achieve the minimum in \eqref{eq:minimization-problem-beta} are characterized by
\begin{align*}
\begin{cases}
&\alpha_t = \beta_t = \id \quad \text{ for all type ``T'' vertices } V_t\\
&\alpha_s = \beta_s = \gamma \quad \text{ for all type ``S'' vertices } V_s\\
&\alpha_i = \beta_i = \id \quad \text{ for all type ``R'' vertices } V_i \text{ situated on a ``TR$\cdots$RT'' arc}\\
&\alpha_j = \beta_j = \gamma \quad \text{ for all type ``R'' vertices } V_j\text{ situated on a ``SR$\cdots$RS'' arc}\\
&\hat 0_p \leq [\alpha_{a_{1}} = \beta_{a_{1}}] \leq \cdots \leq [\alpha_{a_{|a|}} = \beta_{a_{|a|}}] \leq \hat 1_p \quad \text{ for all ``TR$\cdots$RS'' arcs } a \in \mathcal A
\end{cases}
\end{align*}

The asymptotic moments of the marginal are obtained using our main Theorem \ref{thm:moments-network-N}:
\begin{equation}
\lim_{N \to \iy} N^{(k_R + |\mathcal A|)(p-1)}\E \trace\rho_S^p = \prod_{a \in \mathcal A} |\{\text{$|a|$-chains in } NC(p)\}| = \prod_{a \in \mathcal A} FC^{(|a|)}_p = \prod_{a \in \mathcal A} \frac{1}{|a|p +1} \binom{(|a|+1)p}{p}.\end{equation}
The numbers in the right hand side of the equation above are the moments of the following product probability measure:
\begin{equation}\label{eq:limit-measure-cycle}
\pi_S = \times_{a \in \mathcal A} \pi^{(|a|)},
\end{equation}
where $\times$ denotes the classical multiplicative convolution of measures. Recall that if two independent random variables $X$ and $Y$ have respective distributions $\mu$ and $\nu$, then their product $XY$ has distribution $\mu \times \nu$.
\end{proof}

Note that the above result depends only on the partition $\Ptrace = \{S, T\}$ and can be read directly in the graphical representation. The limiting distribution can be inferred from the arcs of type ``TR$\cdots$RS'' of the cycle graph.

\subsection{Exotic graphs}
\label{sec:exotic}

In the previous sections, we discussed graph states which give raise to the following asymptotic spectral distributions:
\begin{enumerate}
\item Maximally mixed states (these correspond to $\delta_1=\pi^{(0)}$),
\item Free Poisson (or Marchenko-Pastur) distribution $\pi^{(1)}$,
\item Fuss-Catalan distribution $\pi^{(s)}$, $s \geq 2$.
\end{enumerate}
All the asymptotic measures we have encountered up to this point are members of the Fuss-Catalan family $\{\pi^{(s)}\}_{s \in \N}$. Moreover, in the preceding section on cycle graphs, we obtained limit eigenvalue distributions which are classical multiplicative convolutions of Fuss-Catalan measures, see equation \eqref{eq:limit-measure-cycle}. In this section, we shall exhibit genuinely new limit asymptotic distributions arising from marginals of random graph states. 

The moments of distributions from the Fuss-Catalan family count the number of chains in the lattice of non-crossing partitions \cite{armstrong}, see also equation \eqref{eq:def-comb-FC}:
\begin{equation}
\int x^p \;\mathrm{d}\pi^{(s)}(x) = |\{\sigma_1, \ldots, \sigma_s \in NC(p) \; | \; \hat 0_p \leq \sigma_1 \leq \cdots \leq \sigma_s \leq \hat 1_p\}|.
\end{equation}

\begin{figure}[htbp]
\centering
\subfigure[]{\label{fig:Hasse-3}\includegraphics{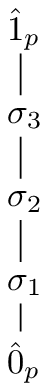}}\qquad\qquad\qquad\qquad
\subfigure[]{\label{fig:Hasse-3-2}\includegraphics{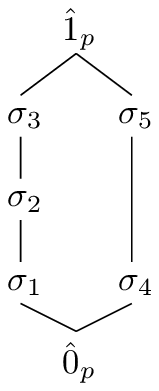}}\qquad\qquad\qquad\qquad
\subfigure[]{\label{fig:Hasse-exotic}\includegraphics{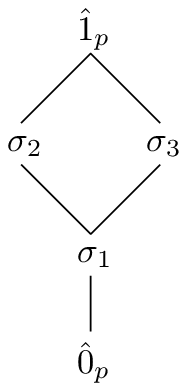}}
\caption{Examples of Hasse diagrams in $NC(p)$: $\pi^{(3)}$, $\pi^{(3)} \times \pi^{(2)}$ and an exotic distribution.}

\end{figure}

In the lattice of non-crossing partitions $NC(p)$, the Hasse diagram of such a chain is presented in Figure \ref{fig:Hasse-3}, for the case $s=3$. Classical multiplicative convolutions of Fuss-Catalan measures correspond to multiple disjoint chains from $\hat 0_p$ and $\hat 1_p$, see Figure \ref{fig:Hasse-3-2} for the Hasse diagram associated to $\pi^{(3)} \times \pi^{(2)}$. The simplest case which can not be described by the Fuss-Catalan statistics and independent products comes from the Hasse diagram in Figure \ref{fig:Hasse-exotic}. Next, we investigate graph state marginals corresponding to this diagram.

\begin{figure}[htbp]
\centering
\subfigure[]{\label{fig:graph-exotic-simple}\includegraphics{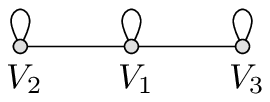}} \qquad
\subfigure[]{\label{fig:graph-exotic}\includegraphics{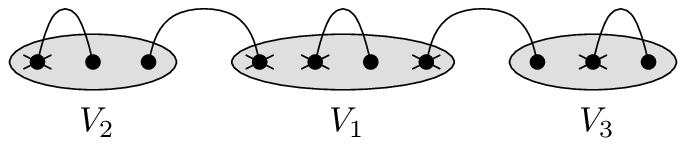}} \\
\subfigure[]{\label{fig:graph-exotic-network}\includegraphics{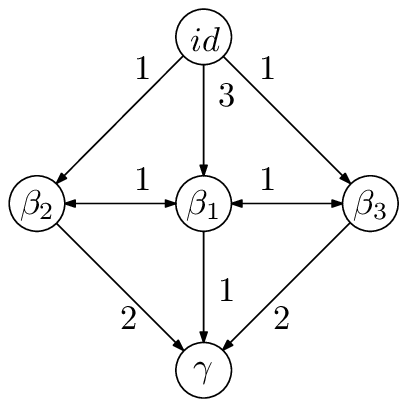}}
\caption{Graph yielding an ``exotic'' limit distribution: (a) the simple form; (b) the marginal yielding the exotic limiting distribution; (c) the associated network.}
\end{figure}

An ad-hoc graph which yields such a Hasse diagram is depicted in Figure \ref{fig:graph-exotic-simple}. The particular marginal of interest $\rho_S = \trace_T \ketbra{\Psi_\text{exotic}}{\Psi_\text{exotic}}$ is drawn in Figure \ref{fig:graph-exotic} with the associated network in Figure \ref{fig:graph-exotic-network}.

The maximum flow problem for the network in Figure \ref{fig:graph-exotic-network} can be easily solved: one can send 5 units of flow from the source $\id$ to the sink $\gamma$ using the following augmenting paths:
\begin{itemize}
\item $\id \to \beta_1 \to \gamma$
\item $\id \to \beta_2 \to \gamma$
\item $\id \to \beta_3 \to \gamma$
\item $\id \to \beta_1 \to  \beta_2 \to \gamma$
\item $\id \to \beta_1 \to  \beta_3 \to \gamma$
\end{itemize}

The residual network is empty and thus the set of permutations $\alpha_i, \beta_i$ achieving the minimum in \eqref{eq:minimization-problem-beta} is given by the conditions $\alpha_i = \beta_i$ for $i=1,2,3$ and by the inequalities of the Hasse diagram in Figure \ref{fig:Hasse-exotic} with $\sigma_i = \beta_i$. The next proposition follows from our main result, Theorem \ref{thm:moments-network-N}.

\begin{proposition}\label{prop:exotic-moments}
The asymptotic moments of the marginal $\rho_S$ are given by
\begin{equation}
\lim_{N \to \iy} \frac{1}{N^5}\E \trace(N^5 \rho_S)^p = |\{\sigma_1, \sigma_2, \sigma_3 \in NC(p) \; | \; \sigma_1 \leq \sigma_2, \; \sigma_1 \leq \sigma_3 \}|.\
\end{equation}
These moments are the moments of the following probability measure
\begin{equation}\label{eq:exotic-measure}
\pi_\text{exotic} = \pi^{(1)} \boxtimes (\pi^{(1)} \times \pi^{(1)}),
\end{equation}
where $\boxtimes$ is the free multiplicative convolution and $\times$ is the classical multiplicative convolution.
\end{proposition}

Notice that the probability measure in equation \eqref{eq:exotic-measure} is genuinely new, and can not be obtain from the Fuss-Catalan family by taking classically independent products. In a similar manner, one can construct graph state marginals with eigenvalue distributions obtained from the free Poisson measures via arbitrary classical and free multiplicative convolutions. Analytical and combinatorial properties of such measures will be investigated in some future work.

\section{Concluding remarks}

In this work we have introduced ensembles of random graph states
and established some of their basic properties. Any graph 
consisting of $k$ vertices and $m$ bonds represents an 
ensemble of pure quantum states, which describe a system
containing $2m$ parts. Each bond represents a
maximally entangled bi-partite pure state, while
each vertex of the graph represents a random coupling between 
the corresponding subsystems.
In the simplest case of the model all subsystems are assumed to be
of the same dimension $N$, which can be treated as a free parameter of the model.
Note that in contrast to generic, structureless random quantum states 
studied by Page \cite{Pa93}, the random states analyzed
in this work do posses certain topological structure determined by the
graph selected.

Dividing the entire multipartite system into two disjoint sets,
one can analyze the typical correlations between these parts.
Technically, one studies the entropy of entanglement
between both subsystems, defined as the von Neumann entropy
of the reduced density matrix.  The key result of this work
consists in developing efficient techniques
which allow one to establish the average 
entropy for an ensemble of states associated with a given graph and
its concrete partition.

In the limit of large $N$ the spectra of random states
obtained by this procedure can be classified.
We have shown for which cases the spectrum can be 
described by the Marchenko-Pastur distribution $\pi^{(1)}$,
also called free Poisson distribution.
This universal distribution describes statistics
of the Wishart matrices $W=G_1G_1^{*}$,
where $G_1$ is a non-hermitian random Ginibre matrix.

In certain graphs states the partial trace leads
to density matrices with spectra described by 
the Fuss-Catalan distribution $\pi^{(s)}$,
studied earlier in \cite{banica-etal}.
These distributions are characteristic of random matrices
with the structure $GG^{*}$,
where $G=\prod_{i=1}^s G_i$ is a product of $s$ 
independent Ginibre matrices. 
Moreover, we identified other examples of the graphs, 
for which the spectra are described by
an even wider class of ``exotic'' distributions.

The model presented here can be generalized in many directions.
Instead of a maximally entangled pure state
$|\Phi^+\rangle=\sum_{i=1}^d \frac{1}{\sqrt{d}} |i\rangle \otimes |i\rangle$
belonging to ${\cal H}_d \otimes {\cal H}_d$,
each edge of the graph could represent a generic random entangled state
$|{\bar \Phi}\rangle=\sum_{i=1}^d \sqrt{p_i} |i\rangle \otimes |i\rangle$,
such that the non-negative numbers $p_i$ sum to unity.
For instance, the random vector $\vec p$ may be generated 
according to the Hilbert-Schmidt measure \cite{SZ04},
which corresponds to taking a random state of size $d^2$.
Such a generalization makes the model more realistic,
but it should not change qualitatively the results obtained,
since the mean entanglement entropy of the corresponding bi-partite system
changes from $\log d$ to $\log d -1/2$.

Any graph analyzed in this work may also represent 
a physical system in a different way.
In this  'dual setup' any bond represents a random unitary matrix
which couples two corresponding subsystems, 
while any vertex with $b$ bonds represents a $b$-particle GHZ-like entangled
state.
 These ensembles, characterized by bi-partite interaction,
are closer related to various physical models of interacting spins.
Analysis of spectral properties of random states 
defined in a the latter procedure is a subject of a following publication.
Observe that for any cycle graph, in which
each vertex has exactly two bonds, both definitions yield 
exactly the same ensembles of random pure states.

We conclude the paper with a rather general remark.
Although the approach presented in this work is not directly related to 
any concrete Hamiltonian model of the physical interaction,
it is capable to describe generic coupling between the subsystems.
Not knowing any details about the kind of the interaction one 
assumes therefore that it can be  mimicked by a 
random unitary matrix.  Averaging over the Haar measure on the unitary group
we obtain rigorous bounds for the average entropy
of entanglement between any two specified fragments of the system.
In the limit of large system size the bounds derived become exact,
and are characteristic of generalized ensembles of random Wishart
matrices. The results obtained for the average entropy are universal in the sense
that they depend on the topology of the graph and its partition,
but do not depend on the kind of the interaction between subsystems.
Establishing a direct link between the results obtained in this work
and the properties of typical matrix product states 
analyzed for concrete physical models  \cite{VC06}
remains as a subject of further investigations.

\bigskip

{\it Acknowledgments: \it}
K.{\.Z}. wishes to thank M. Bo{\.z}ejko for his invitation to Wroc{\l}aw
where this project was initiated, while B.C. was there as
   a Marie Curie Transfer of Knowledge Fellows of the European 
Community's Sixth Framework Programme under contract MTKD-CT-2004-013389.

We also acknowledge the hospitality of the university of Ottawa where the three authors could get together to pursue the project.
K.{\.Z}. is grateful to S.~Braunstein, P.~Horodecki, 
 M.~A.~Nowak, H.-J.~Sommers and F.~Verstraete
for fruitful discussions and e-mail exchange.

K.{\.Z} gratefully acknowledges financial support by the SFB Transregio-12 project,
special grant number DFG-SFB/38/2007 of Polish Ministry of Science and Higher Education,
and  Foundation for Polish Science and European Regional Development Fund
(agreement no MPD/2009/6). 
The research of B.C. and I.N. was partly supported by the NSERC grant 
RGPIN/341303-2007 and the ANR grants Galoisint and Granma.

\end{document}